\PassOptionsToPackage{table}{xcolor}
\pdfoutput=1

\documentclass[dvipsnames]{mytest} 

\usepackage{graphicx}
\usepackage{balance}
\usepackage{tikz}
\usepackage{amsmath}
\usepackage{url}
\usepackage{multirow}
\usepackage{tabularx}
\usepackage{color}
\usepackage{listings}
\usepackage{booktabs} 
\usepackage{makecell}
\usepackage{courier}
\usepackage{caption}
\usepackage{subcaption}
\usepackage{float}
\usepackage{enumitem}

\myTitle{Toward Efficient In-memory Data Analytics on NUMA Systems}
\myAuthors{Puya Memarzia, Suprio Ray, Virendra C Bhavsar}
\myDOI{https://doi.org}
\myVolume{12}
\myNumber{xxx}
\myYear{2019}

\begin{document}
\sloppy
\title{Toward Efficient In-memory Data Analytics\\on NUMA Systems}

\numberofauthors{3}

\author{
\alignauthor
Puya Memarzia\\
\affaddr{Univeristy of New Brunswick}\\
\affaddr{Fredericton, Canada}\\
\email{pmemarzi@unb.ca}
\alignauthor
Suprio Ray\\
\affaddr{Univeristy of New Brunswick}\\
\affaddr{Fredericton, Canada}\\
\email{sray@unb.ca}
\alignauthor
Virendra C Bhavsar\\
\affaddr{Univeristy of New Brunswick}\\
\affaddr{Fredericton, Canada}\\
\email{bhavsar@unb.ca}
}

\maketitle

\begin{abstract}

Data analytics systems commonly utilize in-memory query processing techniques to achieve better throughput and lower latency. Modern computers increasingly rely on Non-Uniform Memory Access (NUMA) architectures in order to achieve scalability. A key drawback of NUMA architectures is that many existing software solutions are not aware of the underlying NUMA topology and thus do not take full advantage of the hardware. Modern operating systems are designed to provide basic support for NUMA systems. However, default system configurations are typically sub-optimal for large data analytics applications. Additionally, achieving NUMA-awareness by rewriting the application from the ground up is not always feasible. 

In this work, we evaluate a variety of strategies that aim to accelerate memory-intensive data analytics workloads on NUMA systems. We analyze the impact of different memory allocators, memory placement strategies, thread placement, and kernel-level load balancing and memory management mechanisms. 
Our findings indicate that the  operating system default configurations can be detrimental to query performance. With extensive experimental evaluation we demonstrate that methodical application of these techniques can be used to obtain significant speedups in four commonplace in-memory data analytics workloads, on three different hardware architectures. Furthermore, we show that these strategies can speed up two popular database systems running a TPC-H workload. 

\category{H.2.4}{Systems}{Query Processing}

\begin{keywords}
NUMA, Memory Allocators, Memory Management, Concurrency, Database Systems, Operating Systems
\end{keywords}
\end{abstract}

\section{Introduction}
\label{sec:intro}

The digital world is producing large volumes of data at increasingly higher rates~\cite{villars2011big,kambatla2014trends,sivarajah2017critical}.  
Data analytics systems are among the key technologies that power the information age. 
The breadth of applications that depend on efficient data processing has grown dramatically. Main memory query processing systems have been increasingly adopted, due to continuous improvements in DRAM capacity and speed, and the growing demands of the data analytics industry \cite{kemper2011hyper}. As the hardware landscape shifts toward greater parallelism and scalability, keeping pace with these changes and maintaining efficiency are a key challenge.


The development of commodity CPU architectures continues to be influenced by various obstacles that hinder the speed and quantity of processing cores that can be packed into a single processor die \cite{asanovic2006landscape}. The power wall motivated the development of multi-core CPUs~\cite{gepner2006multi}, which have become the de facto industry standard. The memory wall~\cite{mckee2004reflections,rogers2009scaling} is a symptom of the growing gap between CPU and memory performance, and the bandwidth starvation of processing cores that share the same memory controller. The demand for greater processing power has pushed the adoption of various decentralized memory controller layouts, which are collectively known as non-uniform memory access (NUMA) architectures. These architectures are widely popular in the server and high performance workstation markets, where they are used for compute-intensive and data-intensive tasks. NUMA architectures are pervasive in multi-socket and in-memory rack-scale systems. Recent developments have led to On-Chip NUMA Architectures (OCNA) that partition the processor's cores into multiple NUMA regions, each with their own dedicated memory controller \cite{molka2015cache, singh20173}. It is clear that the future is NUMA, and that the software stack needs to evolve and keep pace with these changes. Although these advances have opened a path toward greater performance, the burden of efficiently leveraging the hardware has mostly fallen on software developers and system administrators. 

Although a NUMA system's memory is shared among all its processors, the access times to different portions of the memory varies depending on the topology. NUMA systems encompass a wide variety of CPU architectures, topologies, and interconnect technologies. As such, there is no standard for what a NUMA system's topology should look like. Due to the variety of NUMA topologies and applications, fine-tuning the algorithm to a single machine configuration will not necessarily achieve optimal performance on other machines. Given sufficient time and resources, applications could be fine-tuned to the different system configurations that they are deployed on. However, in the real world, this is not always feasible. Therefore, it is desirable to pursue solutions that can improve performance across-the-board, without tuning the code.

In an effort to provide a general solution that speeds up applications on NUMA systems, some researchers have proposed using NUMA schedulers that co-exist with the operating system (OS). These schedulers operate by monitoring running applications in real-time, and managing thread and memory placement \cite{blagodurov2010case, dashti2013traffic, lepers2015thread}. The schedulers make decisions based on memory access patterns, and aim to balance the system load. However, some of these approaches are not architecture or OS independent. For instance, Carrefour~\cite{corbet2012autonuma} needs an AMD CPU based on the K10 architecture, in addition to a modified OS kernel. Moreover, researchers have have argued that  these schedulers may not be beneficial  for  multi-threaded  in-memory query processing~\cite{psaroudakis2015scaling}.
A different approach involves either extensively modifying or completely replacing the operating system. This is done with the goal of providing a custom tailored environment for the application. Some researchers have pursued this direction with the goal of providing an operating system that is more suitable for large database  applications~\cite{giceva2019os,giceva2012towards,giceva2016customized}. Custom operating systems aim to reduce the burden on developers, but their adoption has been limited due to the high pace of advances in both the hardware and software stack. In the past, researchers in the systems community proposed a few new operating systems for multicore architecture, including Corey~\cite{Boyd-Wickizer:2008:Corey}, Barrelfish~\cite{Baumann:2009:TheMultikernel} and fos~\cite{Wentzlaff:2009:FOS}. However, none of them were adopted by the industry. We believe that any custom operating system designed for data analytics will follow the same trajectory. On the other hand, these efforts underscore the need to investigate the impact of system and architectural aspects on query performance.

In recent times, researchers in the database community have started to pay attention to the issues with query performance on  NUMA systems. 
These researchers have favored a more application-oriented approach that involves algorithmic tweaks to the application's source code, particularly, in the context of query processing engines. Among these works some are static solutions that attempted to make query operators NUMA-aware~\cite{schuh2016experimental,wang2015numa}. Others are dynamic solutions that focused on work allocation to threads using work-stealing~\cite{leis2014morsel}, data placement~\cite{kissinger2014eris, Porobic14ATraPos} and task scheduling with adaptive data repartitioning ~\cite{Psaroudakis2016AdaptiveNUMA}.     
These approaches can be costly and time-consuming to implement, and incorporating these solutions to commercial database engines will take time. Regardless, our work is orthogonal to these efforts, as we explore application-agnostic approaches to improve query performance.

Software has been generally slow in adapting to shifts in hardware architecture, such as NUMA. Inefficiencies in the software stack are not always obvious, and the lack of efficient hardware utilization has been easy to overlook in some fields due to a greater focus on multitasking. One common approach is to run multiple tasks (or virtual machines), and give each task a slice of the hardware resources proportional to its needs. This approach is not suitable for data analytics, due to the size of the data, as well as the importance of query throughput and latency. Processing large datasets in main memory data analytics typically calls for a greater emphasis on intra-query parallelism and hardware-awareness.

Main memory data analytics achieve high throughput by leveraging data parallelism on very large sets of memory-resident data, thus diminishing the influence of disk I/O. However, applications that are not NUMA-aware do not fully utilize the hardware's potential \cite{kissinger2014eris}. Furthermore, rewriting the application is not always an option. Solving this problem without extensively modifying the code requires tools and tuning strategies that are application-agnostic. In this work, we evaluate the viability of several key approaches that aim to achieve this. In this context, the impact and role of memory allocators have been under-appreciated and overlooked. We demonstrate that significant  performance gains can be achieved by altering policies that affect thread placement, memory allocation and placement, and load balancing. In particular, we investigate 5 different workloads that prominently feature joins and aggregations, arguably two of the most popular and computationally expensive workloads used in data analytics. Our study covers the following aspects:
\begin{enumerate}[noitemsep,topsep=0pt,parsep=0pt,partopsep=0pt]
\item Dynamic memory allocators (Section~\ref{sec_dynamic_mem_alloc}) \item Thread placement and scheduling (Section~\ref{sec_thread_placement})
\item Memory placement policies (Section~\ref{sec_memory_placement_policies})
\item Operating system configuration: virtual memory page size and NUMA load balancing (Section~\ref{sec_os_config}) 
\end{enumerate}

\begin{figure*}	
	\centering
	\begin{subfigure}[t]{0.375\linewidth}
		\centering
		\includegraphics[height=120pt]{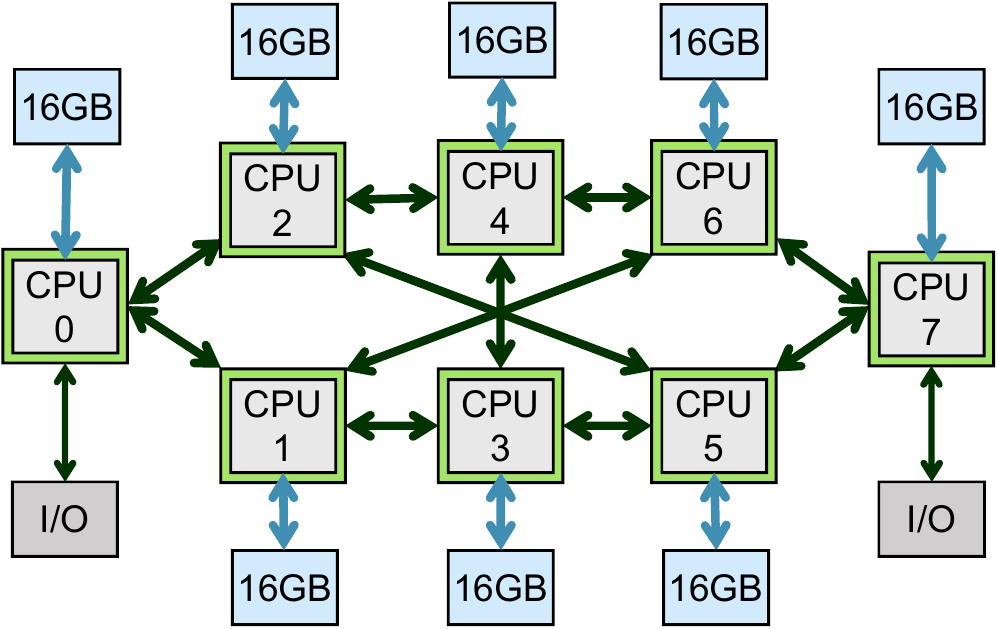}
		\caption{Machine A}\label{fig:top-a}		
	\end{subfigure}
	\hskip -4ex 
	\begin{subfigure}[t]{0.32\linewidth}
		\centering
		\includegraphics[height=120pt]{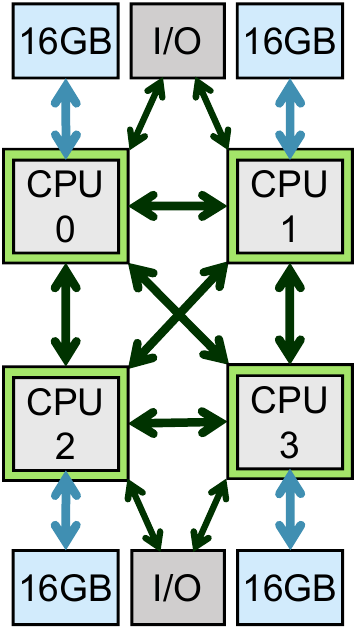}
		\caption{Machine B}\label{fig:top-b}
	\end{subfigure}
	\hskip -10ex
	\begin{subfigure}[t]{0.33\linewidth}
		\centering
		\includegraphics[height=120pt]{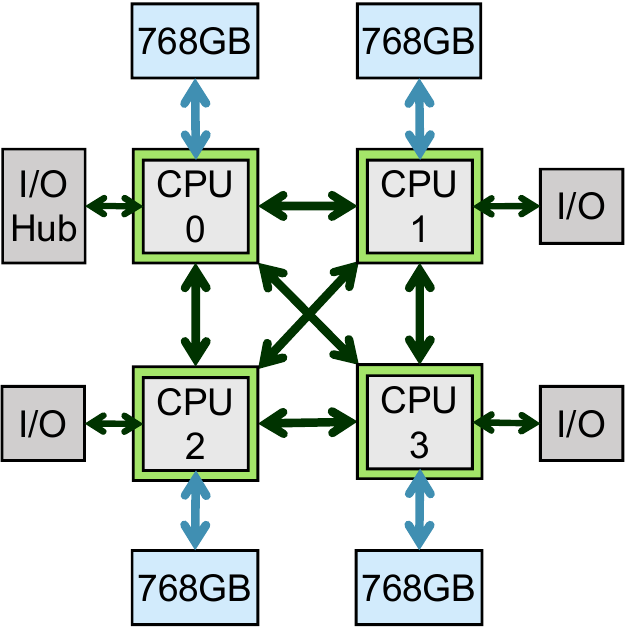}
		\caption{Machine C}\label{fig:top-c}
	\end{subfigure}
	\caption{Machine NUMA Topologies (machine specifications in Table~\ref{tab:machspec})}
	\label{fig:topology}
    \vspace{-6pt} 
\end{figure*}

An important finding from our research is that the default operating system environment can be detrimental to query performance. For instance, the default Linux memory allocator \textit{ptmalloc} can perform poorly compared to other alternatives. Furthermore, with extensive experimental evaluation, we demonstrate that it is possible to systematically utilize \textit{application-agnostic} (or black-box) approaches to obtain speedups on a variety of in-memory data analytics workloads. We show that a hash join workload achieves a $3\times$ speedup on Machine C (see machine topologies in Figure~\ref{fig:topology} and specifications in Table~\ref{tab:machspec}), just from using the \textit{tbbmalloc} memory allocator. This speedup improves to $20\times$ when we utilize the \textit{Interleave} memory placement policy and modify the OS configuration. 
We also show that our findings can carry over to other hardware configurations, by evaluating the experiments on machines with three different hardware architectures and NUMA topologies. Lastly, we show that performance can be improved on two real database systems: MonetDB and PostgreSQL. For example, MonetDB's query latency for the TPC-H workload is reduced by up to 20\% when overriding the memory allocator, and by 43\% by adjusting the operating system configuration.

The main contributions of this paper are as follows:
\begin{itemize}[topsep=3pt,itemsep=2pt,partopsep=2pt, parsep=2pt]
\item Categorization and analysis of the current state-of-the-art strategies to improve application performance on NUMA systems 
\item The first study on NUMA systems (to our knowledge) that explores the \textit{combined impact} of different memory allocators, thread and memory placement policies, and OS-level configurations, on data analytics workloads 
\item Extensive experimental evaluation, including different workloads, machine architectures and topologies, profiling and performance counters, and microbenchmarks 
\item An effective application-agnostic strategic plan to help practitioners speed up memory-intensive applications with minimal code modifications 
\end{itemize}

The remainder of this paper is organized as follows: we provide some background on the problem and elaborate on the workloads in Section~\ref{sec:backgnd}. In Section~\ref{sec:method} we 
discuss the strategies for improving query performance on NUMA systems. We present our setup and experiments in Section~\ref{sec:eval}. We categorize and discuss some of the related work in Section~\ref{sec:related}. Finally, we conclude the paper in Section~\ref{sec:conc}.

\lstset{language=sql}
\lstset{keywordstyle=\color[rgb]{0,0.48,0.60}}
\lstset{morekeywords={MEDIAN}}
\lstset{basicstyle = \ttfamily\small}

\begin{table}[t]
\vspace{-0.2cm}
\small
\setlength\tabcolsep{2.0pt}
\centering
\caption{Experiment Workloads} 
\label{tab:qspecs}
\begin{tabular}{ll} 
\toprule
Workload & SQL Equivalent\\
\midrule
\makecell[l]{\textbf{\textcolor{Blue}{W1)}}Holistic\\Aggregation\\(Hash-based)~\cite{memarzia2019six}} &
\begin{lstlisting}
SELECT groupkey, MEDIAN(val) 
FROM records 
GROUP By groupkey;
\end{lstlisting}\\
\midrule
\makecell[l]{\textbf{\textcolor{Blue}{W2)}}Distributive\\Aggregation\\(Hash-based)~\cite{memarzia2019six}} & 
\begin{lstlisting}
SELECT groupkey, COUNT(val) 
FROM records 
GROUP By groupkey;
\end{lstlisting}\\
\midrule
\makecell[l]{\textbf{\textcolor{Blue}{W3)}}Hash Join~\cite{blanas2011design}\\\\\\} &
\begin{lstlisting}
SELECT * 
FROM table1 INNER JOIN table2 
on table1.jkey = table2.fkey;
\end{lstlisting}\\
\midrule
\makecell[l]{\textbf{\textcolor{Blue}{W4)}}Index Nested\\ Loop Join\\(Different Indexes)\\\cite{leis2016art,mao2012cache,pugh1990skip,skiplist2016}\\\\ } &
\begin{lstlisting}
CREATE INDEX idx_jkey
ON table1 (jkey);
SELECT COUNT(*)
FROM table1 INNER JOIN table2 
on table1.jkey = table2.fkey;
\end{lstlisting}\\
\midrule
\makecell[l]{\textbf{\textcolor{Blue}{W5)}}TPC-H~\cite{council2017tpc}} & 
\makecell[tl]{22 queries that mimic business\\questions on a decision support system.\\Combination of joins and aggregations}\\
\bottomrule
\end{tabular}
\vspace{-0.3cm}
\end{table}

\section{Background}
\label{sec:backgnd}

A NUMA system is divided into several NUMA nodes. Each node consists of one or more processors and their local memory resources. Multiple NUMA nodes are linked together using an interconnect to form a NUMA topology. The topology of our machines is shown in Figure~\ref{fig:topology}. A local memory access involves data that resides on the same node, whereas accessing data on any other node is considered a remote access. Remote data travels over the interconnect, and may need to hop through one or more nodes to reach its destination. Consequently, remote memory access is slower.

In addition to remote memory access, contention is another possible cause of sub-optimal performance on NUMA systems. Due to the memory wall \cite{asanovic2006landscape}, modern CPUs are capable of generating memory requests at a very high rate, which may result in pressure on either the interconnect or the memory controller \cite{dashti2013traffic}. Lastly, the abundance of hardware threads in NUMA systems presents a challenge in terms of scalability, particularly in scenarios with many concurrent memory allocation requests. In Section~\ref{sec:method}, we explore strategies which can be used to mitigate these issues.

\subsection{Experiment Workloads}
\label{sec:workloads}

Our goal is to analyze the effects of NUMA on data analytics workloads, and show effective strategies to gain speedups in these workloads. We have selected five workloads shown in Table \ref{tab:qspecs}, to represent a variety of data operations that are common in data analytics and decision support systems. The implementation of these workloads is described in more detail in Section~\ref{subsec:impl}. We now provide some background on the experiment workloads.

Joins and aggregations are ubiquitous, essential data operations used in many different applications. When use for in-memory query processing, they are notable for stressing the system's memory bandwidth in addition to its capacity. Joins and aggregations are essential components in analytical queries, and are frequently used in popular database benchmarks, such as the TPC-H~\cite{council2017tpc} benchmark. 

A typical aggregation workload involves grouping tuples by a designated grouping column and then applying an aggregate function to each group. Aggregate functions are divided into three categories: distributive, algebraic, and holistic. Distributive functions, such as the \textit{Count} function used in W2 (see Table~\ref{tab:qspecs}), can be decomposed and processed in a distributed manner. This means that the input can be split up, processed, and recombined to produce the final result. Algebraic functions combine two or more distributive functions. For instance, \textit{Average} can be broken down into two distributive functions: \textit{Count} and \textit{Sum}. Holistic aggregate functions, such as the \textit{Median} function used in W1, cannot be decomposed into multiple functions or steps. These aggregate functions do not produce intermediate values, and each output tuple is the result of processing all of the input tuples for its corresponding group. As a result, these aggregate functions are more demanding on the memory system. W3 represents a hash join query. As described in \cite{blanas2011design}, the query joins two tables with a size ratio of 1:16, which is designed to mimic common decision support systems. The join is performed by building a hash table on the smaller table, and probing the larger table for matching keys. W4 is an index nested loop join using the same dataset as W3. The main difference between W3 and W4 is that W3 builds an ad hoc hash table to perform the join, whereas W4 uses a pre-built in-memory index that accelerates lookups to one of the relations. W5 is a database system workload, using the queries and datasets from the TPC-H benchmark~\cite{council2017tpc}. We evaluate W5 on two database systems: MonetDB~\cite{monetdb2018} and PostgreSQL~\cite{postgres2019}.

\section{Improving Query Performance on NUMA Systems}
\label{sec:method}

Achieving good performance on NUMA systems involves careful consideration of thread placement, memory management, and load balancing. We explore application-agnostic strategies that can be applied to the data analytics application in either a black box manner, or with minimal tweaks to the code. Some strategies are exclusive to NUMA systems, whereas others may also yield benefits on uniform memory access (UMA) systems. These strategies consist of: overriding the memory allocator, defining a thread placement and affinity scheme, using a memory placement policy, and changing the operating system configuration.  In this section, we describe these strategies and outline the options used for each one.

\subsection{Dynamic Memory Allocators}
\label{sec_dynamic_mem_alloc}
Dynamic memory allocators are used to track and manage dynamic memory during the lifetime of an application. The performance impact of memory allocators is often overlooked in favor of exploring ways to tweak the application's algorithms. It can be argued that this makes them one of the most under-appreciated system components. Both UMA and NUMA systems can benefit from faster or more efficient memory allocators. NUMA systems typically contain more processing cores, and are particularly sensitive to performance penalties induced by memory access and cache behavior.
Key allocator attributes include allocation speed, fragmentation, and concurrency. Most developers use the default memory allocation functions to allocate or deallocate memory (\textit{malloc}/\textit{new} and \textit{free}/\textit{delete}), and trust that their library will perform these operations efficiently. In recent years, with the growing popularity of multi-threaded applications, there has been a renewed interest in memory allocators, and several alternative allocators have been proposed. Earlier iterations of \textit{malloc} used a single lock which serialized access to the global memory pool. Although recent \textit{malloc} implementations provide support for multi-threaded scalability, there are now several competing memory allocators that aim to reduce multi-threaded contention, and memory consumption overhead.
We evaluate the following allocators: \textit{ptmalloc}, \textit{jemalloc}, \textit{tcmalloc}, \textit{Hoard}, \textit{tbbmalloc}, \textit{mcmalloc}, and \textit{supermalloc}. 

\subsubsection{ptmalloc} 
\textit{ptmalloc} (pthreads malloc) is the memory allocator used in the GNU C Library~\cite{glibc2019} (\textit{glibc}), which is the standard C library in most Linux distributions. It is based on dlmalloc~\cite{dlmalloc2019} (Doug Lea's Malloc). This allocator 
aims to attain a balance between speed, portability, and space-efficiency. \textit{ptmalloc} supports multi-threaded applications by employing multiple mutexes to synchronize and protect access to its data structures. The downside of this approach is the possibility of lock contention on the mutexes. In order to mitigate this issue, \textit{ptmalloc} creates additional regions of memory (arenas) for allocation tasks, whenever contention is detected. A key limitation of \textit{ptmalloc's} arena allocation is that memory can never move between arenas. As of \textit{glibc} version 2.26, which was released in 2017, \textit{ptmalloc} employs a per-thread cache for small allocations. This helps to reduce lock contention by skipping access to the memory arenas when possible. Due to differences in how the machines are configured, we evaluate the versions of \textit{ptmalloc} that shipped with releases 2.27, 2.26, and 2.24 of the \textit{glibc} library. 

\subsubsection{jemalloc}
\textit{jemalloc}~\cite{evans2006scalable} first appeared as a new SMP-aware memory allocator for the FreeBSD operating system, designed by Jason Evans. It was later expanded and adapted for other applications as a general purpose memory allocator. When a thread requests memory from \textit{jemalloc} for the first time, it is assigned a memory allocation arena. For multi-threaded applications, \textit{jemalloc} will assign threads to different arenas in a round-robin fashion. In order to further improve performance, this allocator also uses thread-specific caches, which allows some allocation operations to completely avoid arena synchronization. \textit{jemalloc} divides allocations into three size categories: small (up to 14KB), large (16-3584KB), and huge (4MB+). Lock-free radix trees track allocations across all arenas. \textit{jemalloc} attempts to reduce memory fragmentation by packing allocations into contiguous blocks of memory, and by re-using the first available low address. This approach improves cache locality, but also entails a risk of false sharing, which can hinder performance and must be mitigated by application developers. \textit{jemalloc} provides a solution for this issue by allowing developers to specify cache alignment when allocating memory. To better support NUMA systems, \textit{jemalloc} maintains allocation arenas on a per-CPU basis and associates threads with their parent CPU's arena. We use \textit{jemalloc} version 5.1.0 for our experiments. 

\begin{figure*}
	\centering
	\begin{subfigure}[t]{0.40\linewidth}
		\centering
		\includegraphics[width=\linewidth]{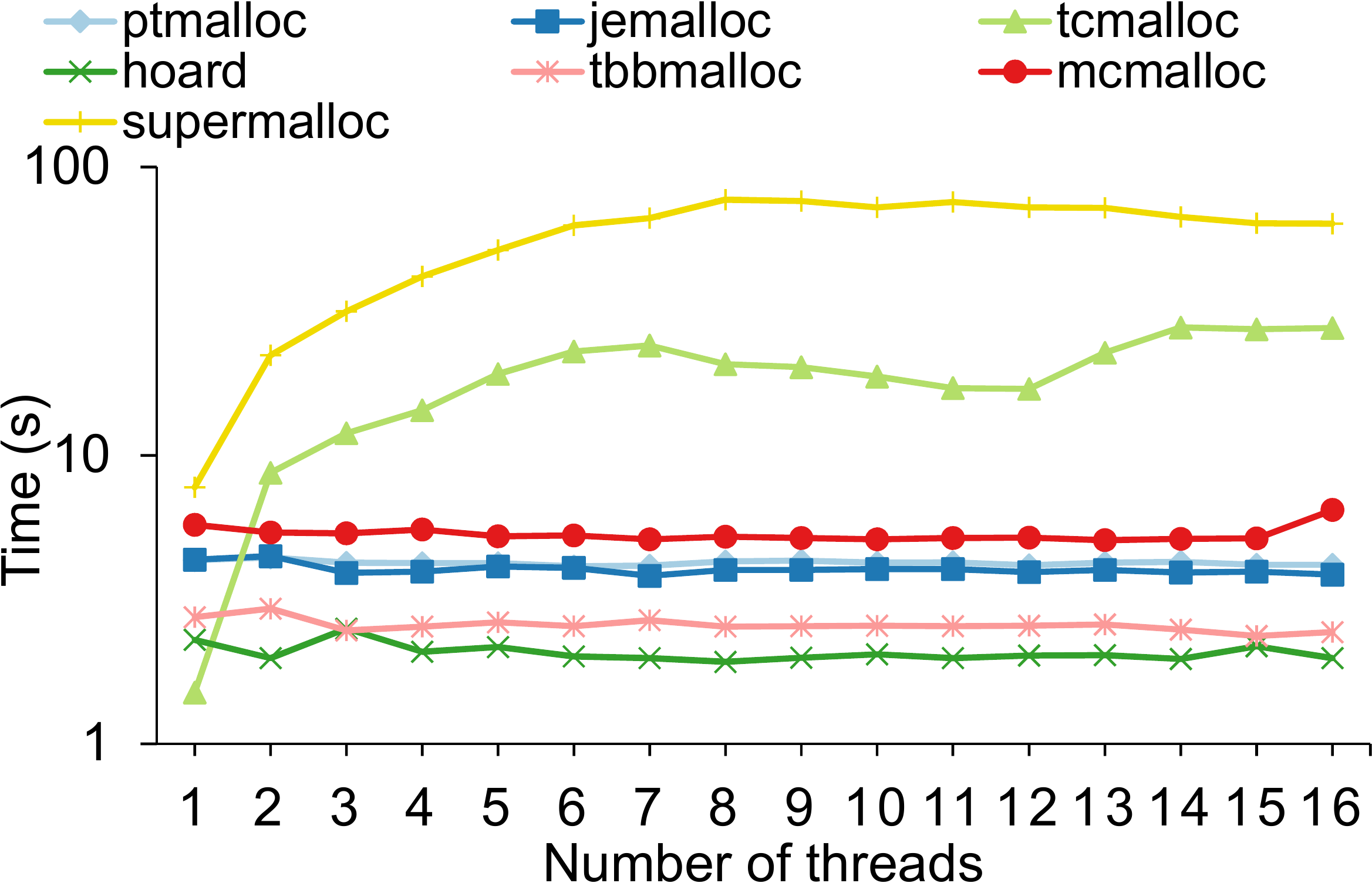}
		\caption{Multi-threaded Scalability}\label{fig:malloc-microbench-time}		
	\end{subfigure}
	\hskip +18ex 
	\begin{subfigure}[t]{0.40\linewidth}
		\centering
		\includegraphics[width=\linewidth]{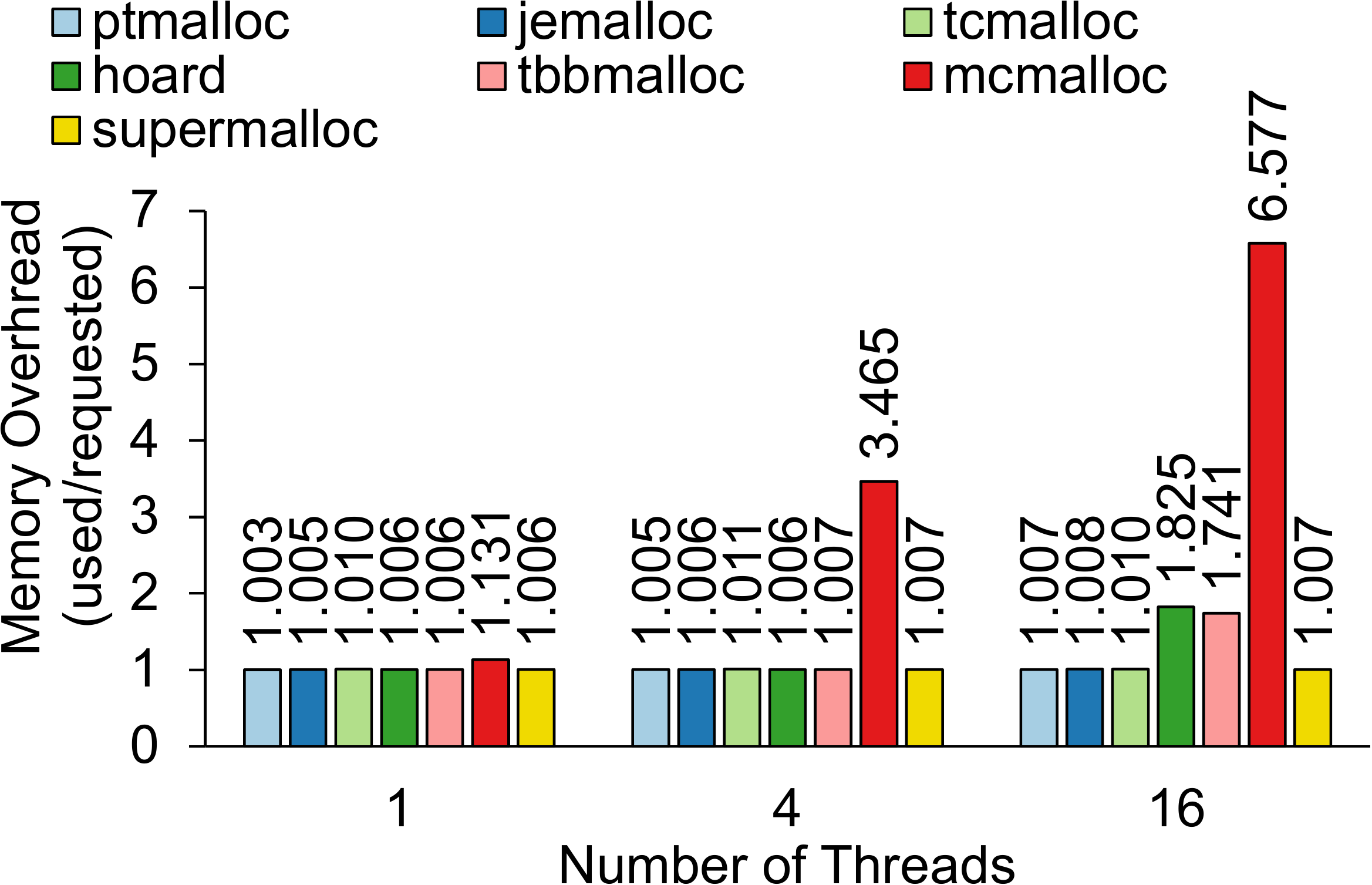}
		\caption{Memory Consumption Overhead}\label{fig:malloc-microbench-overhead}
	\end{subfigure}
	\caption{Memory Allocator Microbenchmark - Machine A}\label{fig:mem-microbench}
	\vspace{-12pt} 
\end{figure*}

\subsubsection{tcmalloc}
The \textit{tcmalloc}~\cite{ghemawat2015tcmalloc} allocator was developed by Google, and is included as part of the \textit{gperftools} library. Its goal is to provide faster memory allocations in memory-intensive multi-threaded applications. \textit{tcmalloc} divides allocations into two categories: large allocations and small allocations. Large allocations use a central heap that is organized into contiguous groups of pages called ``spans''. Each span is designed to fit multiple allocations (regions) of a particular size class. Since all the regions in a span are of the same size, only one metadata header is maintained for each span. However, allocations from a size class cannot be allocated inside spans for other classes. As a result, applications that use many different classes may waste memory due to inefficient utilization of the memory spans. The central heap uses fine-grained locking on a per-class basis. As a result, two threads requesting memory from the central heap can do so concurrently, as long as their requests fall in different class categories. Small allocations are served by private thread-local caches and do not require any locking. We use the version of \textit{tcmalloc} included in \textit{gperftools} 2.7.

\subsubsection{Hoard}
\textit{Hoard}~\cite{berger2000hoard} is a standalone cross-platform allocator replacement designed specifically for multi-threaded applications. \textit{Hoard}'s main design goals are to provide memory efficiency, reduce allocation contention, and prevent false sharing. At its core, \textit{Hoard} consists of a global heap (the ``hoard'') that is protected by a lock and accessible by all threads, as well as per-thread heaps that are mapped to each thread using a hash function. The allocator counts the number of times that a thread has acquired the global heap lock in order to decide if contention is occurring. \textit{Hoard} also employs heuristics to detect temporal locality, and uses this information to fill cache lines with objects that were allocated by the same thread, thus avoiding false sharing. Recent updates to \textit{Hoard} have increased the size of the per-thread heap, and reduced the heap layer overhead. We evaluate \textit{Hoard} version 3.13 in our experiments.

\subsubsection{tbbmalloc}
The \textit{tbbmalloc}~\cite{kukanov2007foundations} allocator is included as part of the \textit{Intel Thread Building Blocks} (\textit{TBB}) library~\cite{kim2011multicore}. It is based on some of the concepts and ideas outlined in their prior work on  McRT-Malloc~\cite{hudson2006mcrt}. This allocator pursues better performance and scalability for multi-threaded applications, and generally considers increased memory consumption as an acceptable tradeoff. In response to the issues with memory footprint, \textit{TBB} 4.2 update 1 (released in 2014) allowed developers to set a soft limit on the allocator's memory consumption. Reaching this limit triggers the allocator's internal buffers to free their memory. Allocations in \textit{tbbmalloc} are supported by per-thread memory pools. If the allocating thread is the owner of the target memory pool, no locking is required. If the target pool belongs to a different thread then the request is placed in a synchronized linked list, and the owner of the pool will allocate the object. We used version 2019 Update 4 of the \textit{TBB} library for our experiments. 

\subsubsection{supermalloc}
\textit{supermalloc}~\cite{kuszmaul2015supermalloc} is a \textit{malloc} replacement that synchronizes concurrent memory allocation requests using hardware transactional memory (HTM) if available, and falls back to \textit{pthread} mutexes if HTM is not available. It prefetches all necessary data while waiting to acquire a lock in order to minimize the amount of time spent in the critical section. \textit{supermalloc} uses homogeneous chunks of objects for allocations smaller than 1MB, and supports larger objects using operating system primitives. In order to reduce conflicts between different class sizes, each class is a prime multiple of the cache line size. Given a pointer to an object, its corresponding chunk is tracked using a look up table. The chunk table is implemented as a large 512MB array, but the allocator takes advantage of the fact that most of its virtual memory will not be committed to physical memory by the operating system. For our experiments, we use the latest publicly released source code, which was last updated in October 2017. 

\subsubsection{mcmalloc}
\textit{mcmalloc}~\cite{umayabara2017mcmalloc} focuses on mitigating multi-threaded lock contention by reducing calls to kernel space, dynamically adjusting the memory pool structures, and using fine-grained locking. Similar to other allocators, it uses a global and local (per-thread) memory pool layout. \textit{mcmalloc} monitors allocation requests, and dynamically splits its global memory pool into two categories: frequently used memory chunk sizes, and infrequently used memory chunk sizes. Dedicated homogeneous memory pools are created to support frequently used chunk sizes. Infrequent memory chunk sizes are handled using size-segregated memory pools. \textit{mcmalloc} reduces system calls by batching multiple chunk allocations together, and by not returning memory to the OS when \textit{free} is called. We use the latest \textit{mcmalloc} source code, which was updated in March 2018. 

\subsubsection{Memory Allocator Microbenchmark}
\label{subsec:microbench}
We now describe a multi-threaded microbenchmark that we use to gain insight on the relative performance of these memory allocators. The goal of the microbenchmark is to answer the question: how well do these allocators scale up on a NUMA machine? This experiment simulates a memory-intensive workload with multiple threads utilizing the allocator at the same time. Each thread completes 100 million memory operations, consisting of allocating memory and writing to it, or reading an existing item and then  deallocating it. The distribution of allocation sizes is inversely proportional to the size class (smaller allocations are more frequent). We use two metrics to compare the allocators: execution time, and memory allocation overhead. The execution time gives an idea of how fast an allocator is, as well as its efficiency when being used in a NUMA system by concurrent threads. In Figure~\ref{fig:malloc-microbench-time}, we vary the number of threads in order to see how each allocator behaves under contention. The results show that \textit{tcmalloc} provides the fastest single-threaded performance, but falls behind as the number of threads is increased. \textit{Hoard} and \textit{tbbmalloc} show good scalability, and outperform the other allocators by a considerable margin. In Figure~\ref{fig:malloc-microbench-overhead}, we show each  allocator's overhead. This is calculated by measuring the amount of memory allocated by the operating system (as maximum resident set size), and dividing it by the amount of memory that was requested by the microbenchmark. This experiment shows considerably higher memory overhead for \textit{mcmalloc} as the number of threads increases. \textit{Hoard} and \textit{tbbmalloc} are slightly more memory hungry than the other allocators. Based on these results, we omit \textit{supermalloc} and \textit{mcmalloc} from subsequent experiments, due to their poor performance in terms of scalability and memory overhead respectively.

\begin{figure}
\begin{minipage}[t]{0.47\linewidth}
    \includegraphics[width=\linewidth]{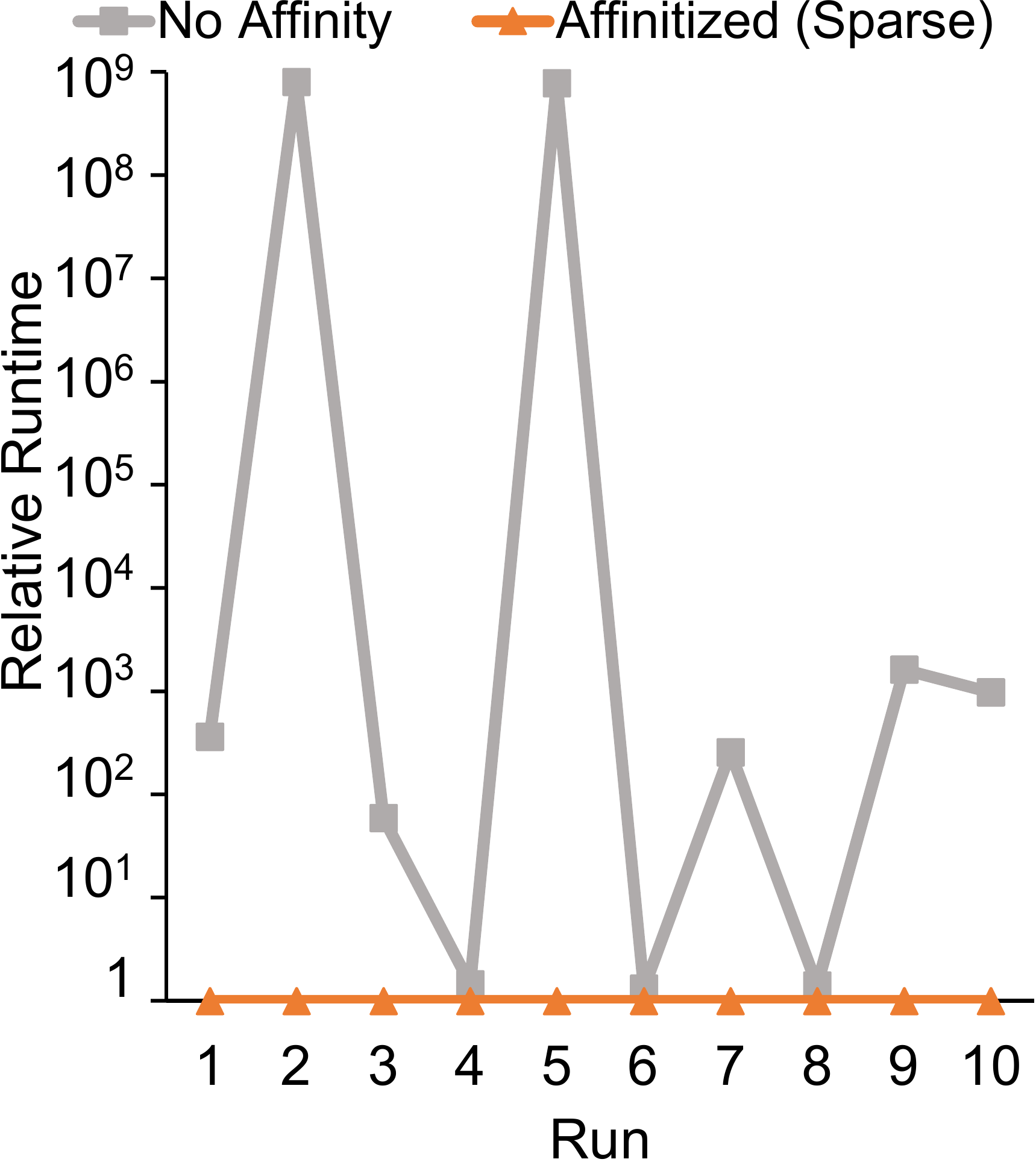}
    \vspace{+0.05cm}
    \caption{Multiple runs of the holistic aggregation workload (W1) - affinitized threads versus default operating system scheduling - Machine A (16 threads)} 
    \label{fig:defaultvsaffinitized}
\end{minipage}
    \hfill
\begin{minipage}[t]{0.465\linewidth}
    \includegraphics[width=\linewidth]{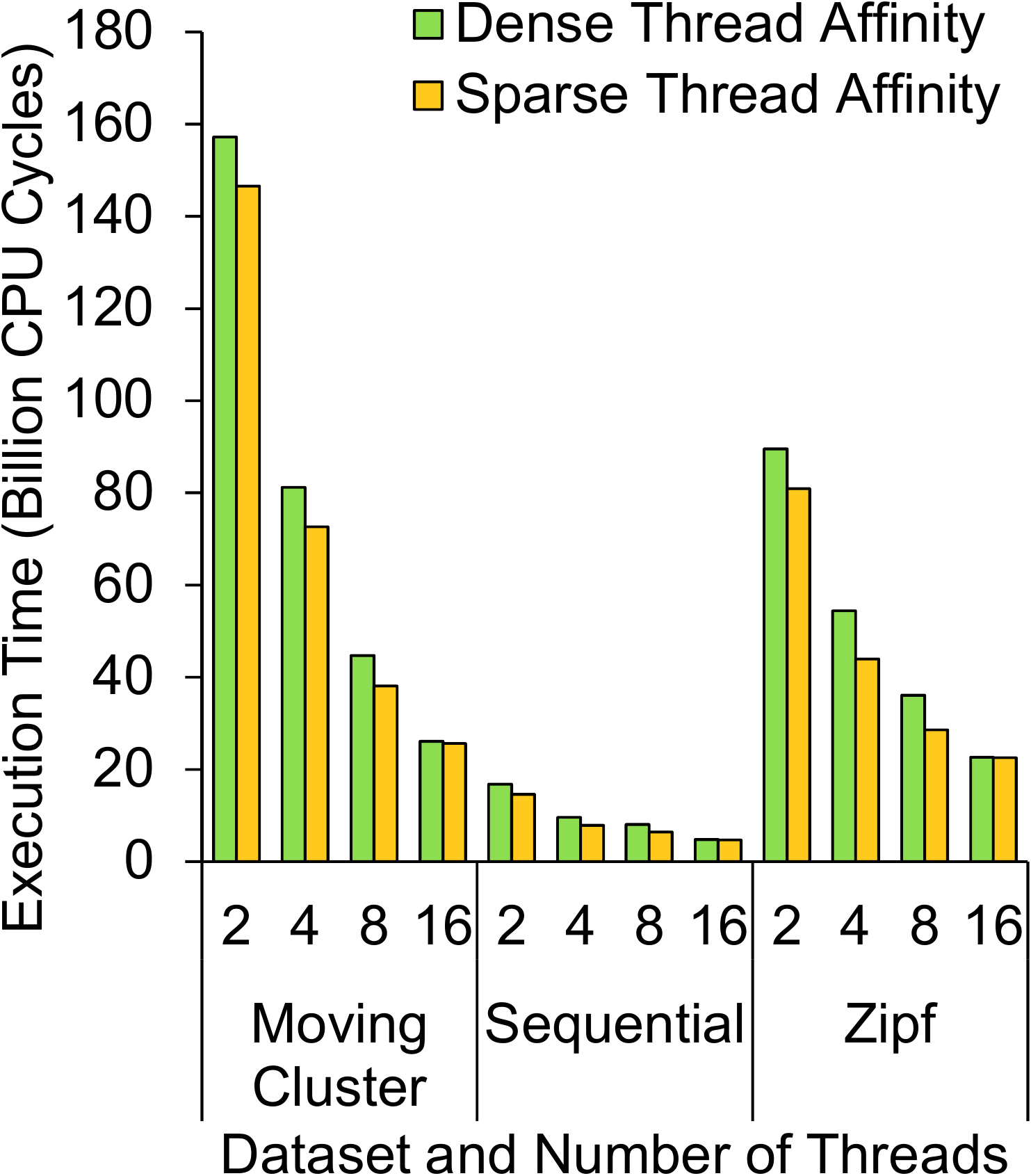}
    \vspace{+0.05cm}
    \caption{Comparison of two thread affinitization strategies - Holistic Aggregation Workload (W1) - Machine A}
    \label{fig:densevssparse}
\end{minipage}
\end{figure}

\subsection{Thread Placement and Scheduling} 
\label{sec_thread_placement}
Defining an efficient thread placement strategy is a well-known and essential step toward obtaining better performance on NUMA systems. By default, the kernel thread scheduler is free to migrate threads created by the program between all available processors. The reasons for doing so include power efficiency and balancing the heat output of different processors. This behavior is not ideal for large data analytics applications, and may result in significantly reduced query throughput. The thread migrations slow down the program due to cache invalidation, as well as a likelihood of moving threads away from their local data. The combination of cache invalidation, loss of locality, and non-deterministic behavior of the OS scheduler can result in wild performance fluctuations (as depicted in Figure \ref{fig:defaultvsaffinitized}). Binding threads to processor cores can solve this issue by preventing the OS from migrating threads. However, deciding how to place the threads requires careful consideration of the topology, as well as the software environment.

\begin{table}
\setlength\tabcolsep{5.5pt}
\centering
\caption{Profiling holistic aggregation workload (W1) - Machine A (16 threads) - Impact of thread affinity - \textbf{Default} (managed by operating system) vs  \textbf{Modified}  (\textit{Sparse} thread placement)}
\label{tab:profaffinity}
\begin{tabular}{lrrr} 
\toprule
\textbf{Metric} &\textbf{Default} &  \textbf{Modified} & \textbf{Diff}\\
\midrule
Thread Migrations & 33196 &  16 & -99.95\%\\
\midrule
Cache Misses & 1450M  & 972M & -32.95\%\\
\midrule
\makecell[l]{Local Memory\\Access} & 367M &  374M & +2.06\%\\
\midrule
\makecell[l]{Remote Memory\\Access} & 159M &  108M & -31.95\%\\
\midrule
Local Access Ratio & 0.70 &  0.78 & +10.77\%\\
\bottomrule
\end{tabular}
\vspace{-0.2cm}
\end{table}

A thread placement strategy details the manner in which threads are assigned to processors. We explore two strategies for assigning thread affinity: \textit{\textbf{Dense}} and \textit{\textbf{Sparse}}. A \textit{Dense} thread placement involves packing threads in as few processors as possible. The idea behind this approach is to minimize remote access distance and maximize resource sharing. In contrast, the \textit{Sparse} strategy attempts to maximize memory bandwidth utilization by spreading the threads out among the processors. There are a variety of ways to implement and manage thread placement, depending on the level of access to the source code and the library used to provide multithreading. Applications built on OpenMP can use the \textit{OMP\_PROC\_BIND} and \textit{OMP\_PLACES} environment variables in order to fine-tune thread placement at runtime. If none of the above options are feasible, the \textit{numactl} tool can be used to bind the application process to a specific set of processors, but does not prevent migrations within the set. 

To demonstrate the impact of affinitization, we evaluate workload W1 from Table~\ref{tab:qspecs}, using Machine A shown in Figure~\ref{fig:topology}. The workload involves building a hash table with key-value pairs taken from a moving cluster distribution. Figure~\ref{fig:defaultvsaffinitized} depicts 10 
consecutive runs of this workload. The runtime number of the default configuration (\textit{no affinity}) is expressed in relation to the \textit{affinitized} configuration. The results highlight the inconsistency of the operating system's default behavior. In the best case, the \textit{affinitized} configuration is several orders of magnitude faster, and the worst case runtime is still around $27\%$ faster. In order to gain a better understanding of how each configuration affects the workload, we use the \textit{perf} tool to measure several key metrics. The results depicted in Table~\ref{tab:profaffinity}, show that the operating system is migrating threads many times during the course of a workload. The \textit{sparse} affinity configuration prevents migration-induced cache invalidation, which in turn reduces cache misses. Furthermore, the stabilized thread placement increases the ratio of memory accesses that are satisfied by local memory, resulting in more bandwidth.

In Figure \ref{fig:densevssparse} we evaluate the \textit{sparse} and \textit{dense} thread affinity strategies on workload W1, and vary the number of threads. We also vary the dataset (see Section~\ref{subsec:impl}) in order to ensure that the distribution of the data records is not the defining factor. The goal of this experiment is to determine if threads benefit from being on the same NUMA node against utilizing a greater number of the system's memory controllers. The \textit{sparse} policy achieves better performance when the workload is not using all available hardware threads. This is due to the threads having access to additional memory bandwidth, which plays a major role in memory-intensive workloads. When all hardware threads are occupied, the two policies perform almost identically. Henceforth, we use the \textit{sparse} configuration (when applicable) for all our experiments. 

\subsection{Memory Placement Policies}
\label{sec_memory_placement_policies}
Memory pages are not always accessed from the same threads that allocated them. Memory placement policies are used to control the location of memory pages in relation to the NUMA topology. As a general rule of thumb, data should be on the same node as the thread that processes it, and sharing should be minimized. However, too much consolidation can lead to congestion of the interconnects, and contention on the memory controllers. The \textit{numactl} tool applies a memory placement policy to a process, which is then inherited by all its children (threads). We evaluate the following policies: \textit{First Touch}, \textit{Interleave}, \textit{Localalloc}, and \textit{Preferred}. We also use hardware counters to measure the ratio of local to total (local+remote) memory accesses.

Modern Linux systems employ a memory placement policy called \textit{\textbf{First Touch}}. In \textit{First Touch}, each memory page is allocated to the first node that performs a read or write operation on it. If the selected node does not have sufficient free memory, an adjacent node is used. This is the most popular memory placement policy, and represents the default configuration for most Linux distributions. \textit{\textbf{Interleave}} places memory pages on all NUMA nodes in a round-robin fashion. In some prior works, memory interleaving was used to spread a shared hash table across all available NUMA nodes \cite{balkesen2013main, LangLA0K13, leis2014morsel}. In \textit{\textbf{Localalloc}}, the memory pages are placed on the same NUMA node as the thread performing the allocation. The \textit{\textbf{Preferred$x$}} policy places all newly allocated memory pages on node $x$. It will use other nodes for allocation only when  node $x$ has run out of free space and cannot fulfill the allocation.

\subsection{Operating System Configuration}
\label{sec_os_config}
In this section, we outline two key operating system mechanisms that affect NUMA applications: Virtual Memory Page Size (Transparent Hugepages), and Load Balancing Schedulers (AutoNUMA). These mechanisms are enabled out-of-the-box on most Linux distributions. 

\subsubsection{Virtual Memory Page Size} 

Operating system memory management works at the virtual page level. Pages represent chunks of memory, and their size determines the granularity of which memory is tracked and managed. 
Most Linux systems use a default memory page size of 4KB in order to minimize wasted space. The CPU's TLB caches can only hold a limited number of page entries. When the page size is larger, each TLB entry spans a greater memory area. Although the TLB capacity is even smaller for large entries, the total volume of cached memory space is increased. As a result, larger page sizes may reduce the occurrence of TLB misses. Transparent Hugepages (THP) is an abstraction layer that automates the process of creating large memory pages from smaller pages. Some prior works have found that larger memory pages can improve query runtimes by reducing TLB misses~\cite{leis2014morsel,schuh2016experimental}. These findings are not universal however, as several product documentations recommend disabling THP, including the Red Hat Performance Tuning Guide~\cite{redhat2018thp}, Oracle~\cite{oracle2019}, Redis~\cite{redis2019}, and MongoDB~\cite{mongodb2019}. Other database systems, such as VoltDB~\cite{stonebraker2013voltdb} will refuse to start until THP has been disabled. Reasons cited include incompatibilities with existing memory management framework, increased memory consumption, and additional swapping latency. The hardware architecture also plays an important role, as the size of the TLB cache varies between different CPU architecture. On Linux machines, control over the page size is provided by the \textit{Transparent Hugepages (THP)} library. We evaluate the effect of using 4KB (default) and 2MB memory pages. 

\subsubsection{Automatic NUMA Load Balancing} 
There have been several projects to develop NUMA-aware schedulers that facilitate automatic load balancing. Among these projects, \textit{Dino} and \textit{AsymSched} do not provide any source code, and \textit{Numad} is designed for multi-process load balancing. \textit{Carrefour}~\cite{dashti2013traffic} provides public source code, but requires an AMD CPU based on the K10 architecture (with instruction-based sampling), as well as a modified operating system kernel. Consequently, we opted to evaluate the \textit{AutoNUMA} scheduler, which is open-source and supports all hardware architectures. \textit{AutoNUMA} was initially developed by Red Hat and later on merged with the Linux kernel. It attempts to maximize data and thread co-location by migrating memory pages and threads. \textit{AutoNUMA} has two key limitations: 1) workloads that utilize data sharing can be mishandled as memory pages may be continuously unnecessarily migrated between nodes 2) it does not factor in the cost of migration or contention, and thus aims to improve locality at any cost. \textit{AutoNUMA} has received continuous updates, and is considered to be one of the most well-rounded kernel-based NUMA schedulers. We use the \textit{numa\_balancing} kernel parameter to enable or disable this NUMA scheduler.
\section{Evaluation}
\label{sec:eval}

\begin{table}[t]
\setlength\tabcolsep{3.0pt}
\centering
\caption{Machine Specifications}
\label{tab:machspec}
\begin{tabular}{cccc}
\toprule
System & \textbf{Machine A} & \textbf{Machine B} & \textbf{Machine C}\\
\midrule
\makecell{CPUs/\\Model} & \makecell{8$\times$Opteron\\8220} & \makecell{4$\times$Xeon\\E7520} & \makecell{4$\times$Xeon\\E7-4850 v4} \\
\midrule
\makecell{CPU\\Frequency} & 2.8GHz & 2.1GHz & 2.1GHz\\
\midrule
\makecell{Architecture} & \makecell{AMD\\Santa Rosa} & \makecell{Intel\\Nehalem} & \makecell{Intel\\Broadwell}\\ 
\midrule
\makecell{Physical/\\Logical Cores} & 16/16 & 16/32 & 32/64\\
\midrule
\makecell{Last Level\\Cache} & 2MB & 18MB & 40MB\\
\midrule
\makecell{4KB TLB\\Capacity} & \makecell[l]{L1:$32\times$4KB\\L2:$512\times$4KB} & \makecell[l]{L1:$64\times$4KB\\L2:$512\times$4KB} & \makecell[l]{L1:$64\times$4KB\\L2:$1536\times$4KB}\\
\midrule
\makecell{2MB TLB\\Capacity} & 
\makecell[c]{L1:$8\times$2MB\\-} & \makecell[c]{L1:$32\times$2MB\\-} & \makecell[l]{L1:$32\times$2MB\\L2:$1536\times$2MB}\\
\midrule
\makecell{NUMA\\Nodes} & 8 & 4 & 4\\
\midrule
\makecell{NUMA\\Topology} & \makecell{Twisted\\Ladder} & \makecell{Fully\\Connected} & \makecell{Fully\\Connected}\\
\midrule
\makecell{Relative\\NUMA Node\\Memory\\Latency} & \makecell[r]{Local: 1.0\\1 hop: 1.2\\2 hop: 1.4\\3 hop: 1.6} & \makecell[r]{Local: 1.0\\1 hop: 1.1\\\\\\ } & \makecell[r]{Local: 1.0\\1 hop: 2.1\\\\\\ }\\
\midrule
\makecell{Interconnect\\Bandwidth} & \makecell{2GT/s} & \makecell{4.8GT/s} & \makecell{8GT/s}\\
\midrule
\makecell{Memory\\Capacity} & \makecell{16GB/node\\128GB Total} & \makecell{16GB/node\\64GB Total} & \makecell{768GB/node\\3TB Total}\\
\midrule
\makecell{Memory\\Clock} & \makecell{800MHz} & \makecell{1600MHz} & \makecell{2400MHz}\\
\midrule
\makecell{Operating\\System} & \makecell{Ubuntu\\16.04} & \makecell{Ubuntu\\18.04} & \makecell{CentOS\\7.5} \\
\midrule
\makecell{Linux\\Kernel} & \makecell{4.4\\x86\_64} & \makecell{4.15\\x86\_64} & \makecell{3.10\\x86\_64} \\
\midrule
\makecell{C++ library\\(\textit{glibc})} & 2.26 & 2.27 & 2.24 \\
\bottomrule
\end{tabular}
\vspace{-0.3cm}
\end{table}

In this section, we describe our setup, and evaluate the effectiveness of our techniques. In Section~\ref{sec:setup} we outline the specifications of our machines, as well as the software configuration. We begin by analyzing the impact of the operating system configuration in Section~\ref{subsec:results-anthp}. In Section~\ref{subsec:results-tpch} we evaluate these techniques on database engines running TPC-H queries. We explore the effects of overriding the default system memory allocator in Section~\ref{subsec:results-malloc}. Finally, we summarize our findings in Section~\ref{sec:summary}.

\subsection{Experimental Setup}
\label{sec:setup}

\begin{table}[t]
\setlength\tabcolsep{7.0pt}
\centering
\caption{Experiment Parameters (bolded values are used as defaults)}
\label{tab:expar}
\begin{tabular}{>{\raggedright\arraybackslash}p{2.6cm}>{\raggedright\arraybackslash}p{4.8cm}}
\toprule
\textbf{Parameter} &\textbf{Values} \\
\midrule
Experiment Workload & \makecell[tl]{\textbf{W1)} Holistic Aggregation~\cite{memarzia2019six}\\W2) Distributive Aggregation\hspace{0.8pt}\cite{memarzia2019six}\\W3) Hash Join~\cite{blanas2011design}\\W4) Index Nested Loop Join~\cite{leis2016art}\\W5) TPC-H Query~\cite{council2017tpc}}\\
\midrule
Thread Placement Policy & None (operating system is free to migrate threads), \textbf{Sparse}, Dense \\
\midrule
Memory Placement Policy & \textbf{First Touch}, Interleaved, Localalloc, Preferred$x$\\
\midrule
Memory Allocator & \textbf{ptmalloc}, jemalloc, tcmalloc, Hoard, tbbmalloc\\ 
\midrule
Dataset Distribution & \textbf{Moving Cluster}, Sequential, Zipf, TPC-H Dataset\\
\midrule
Operating System Configuration & \textbf{AutoNUMA on}/off, \textbf{Transparent Hugepages (THP) on}/off\\
\midrule
Hardware System & \textbf{Machine A}, Machine B, Machine C\\
\bottomrule
\end{tabular}
\vspace{-0.4cm}
\end{table}

We run our experiments on three machines based on completely different architectures. This is done to ensure that our findings are not biased by a particular system's characteristics. The NUMA topologies of these machines are depicted in Figure~\ref{fig:topology} and their specifications are outlined in Table~\ref{tab:machspec}. We used \textit{LIKWID}~\cite{likwid2010} to measure each system's relative memory access latencies, and the remainder of the specifications were obtained from product pages, spec sheets, and Linux system queries. Now we outline some of the key hardware specifications for each machine. Machine A is an eight socket AMD-based server, with a total of 128GB of memory. As the only machine with eight NUMA nodes, machine A provides us with an opportunity to study NUMA effects on a larger scale. The twisted ladder topology shown in Figure~\ref{fig:top-a} is designed to minimize inter-node latency with three HyperTransport interconnect links per node. As a result, Machine A has three categories of memory access latencies, depending on number of hops required to get from the origin to the destination of the memory access. Each node contains an AMD Opteron 8220 CPU running at 2.8GHz and 16GB of memory. Each of the Opteron 8220's cores feature a 128KB L1 cache, and a 2MB L2 cache. Machine B is a quad-socket Intel server with four NUMA nodes and a total memory capacity of 64GB. The NUMA nodes are fully connected, and each node consists of an Intel Xeon E7520 CPU running at 1.87GHz, and 16GB of memory. Each core in the Xeon E7520 features a 256KB L1 and 1MB L2 cache, and an 18MB L3 cache that is shared between all cores. Lastly, Machine C contains four sockets populated with Intel Xeon E7-4850 v4 processors. Each processor constitutes a NUMA node with 768MB of memory, providing a total system memory capacity of 3TB. The NUMA nodes of this machine are fully connected. Each processor is equipped with 40MB of L3 cache that is shared between all cores, and each core features 256KB of L2 cache and 64KB of L1 cache. 

The code for all our experiments is written in C++ and compiled using GCC 7.3.0 with the -O3 and -march=native flags. Likewise, all dynamic memory allocators are synchronized to the same versions, and compiled from source on each machine. Machines B and C are owned and maintained by external parties, and are based on different Linux distributions. Unless otherwise noted, all experiments are configured to utilize all available hardware threads.

\begin{figure*}[ht]
	\centering
	\begin{subfigure}[t]{0.205\linewidth}
		\centering
		\includegraphics[width=\linewidth]{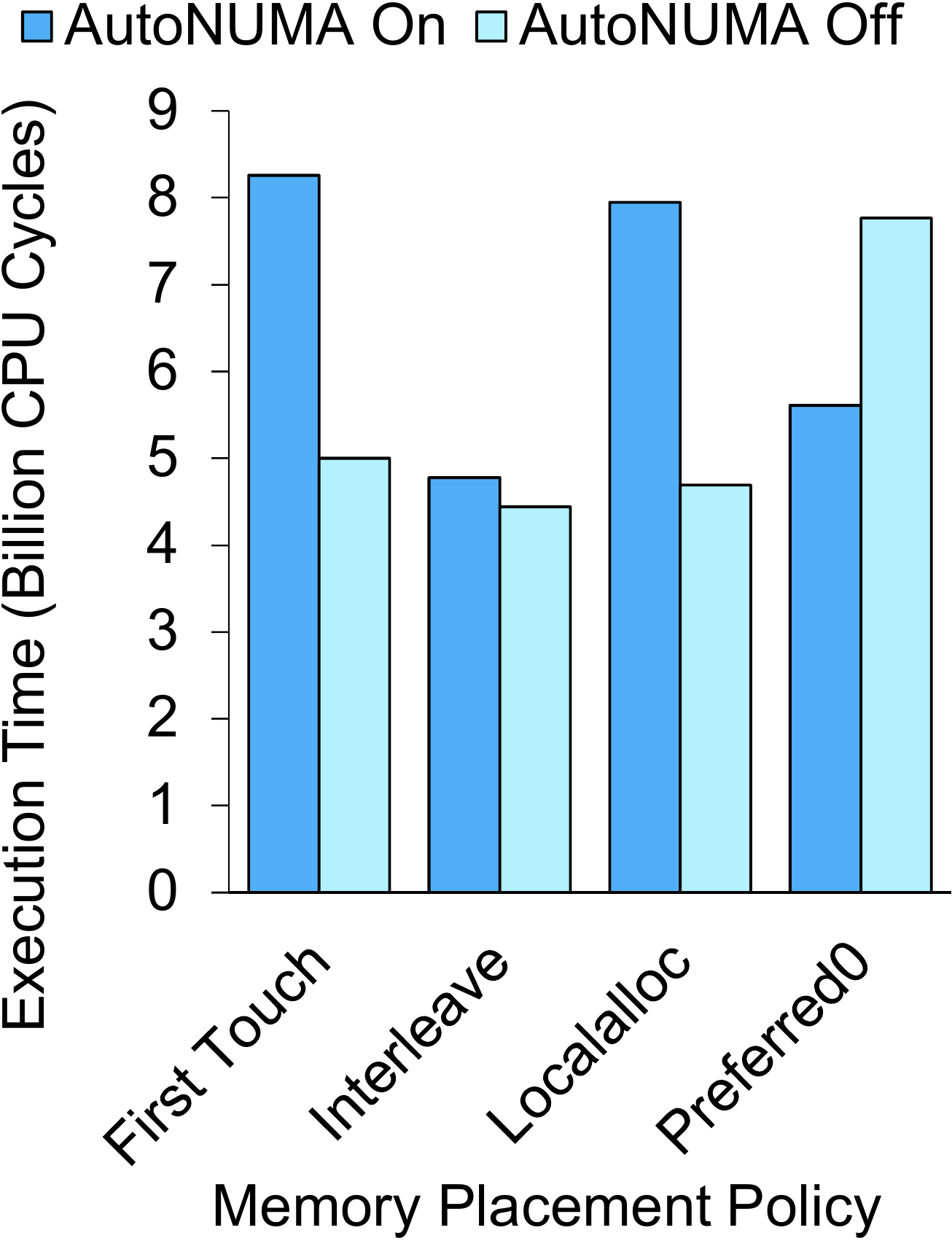}
		\caption{AutoNUMA effect on execution time - Machine A}\label{fig:osconfig-autonuma-runtime}		
	\end{subfigure}
	\hfill
	\begin{subfigure}[t]{0.205\linewidth}
		\centering
		\includegraphics[width=\linewidth]{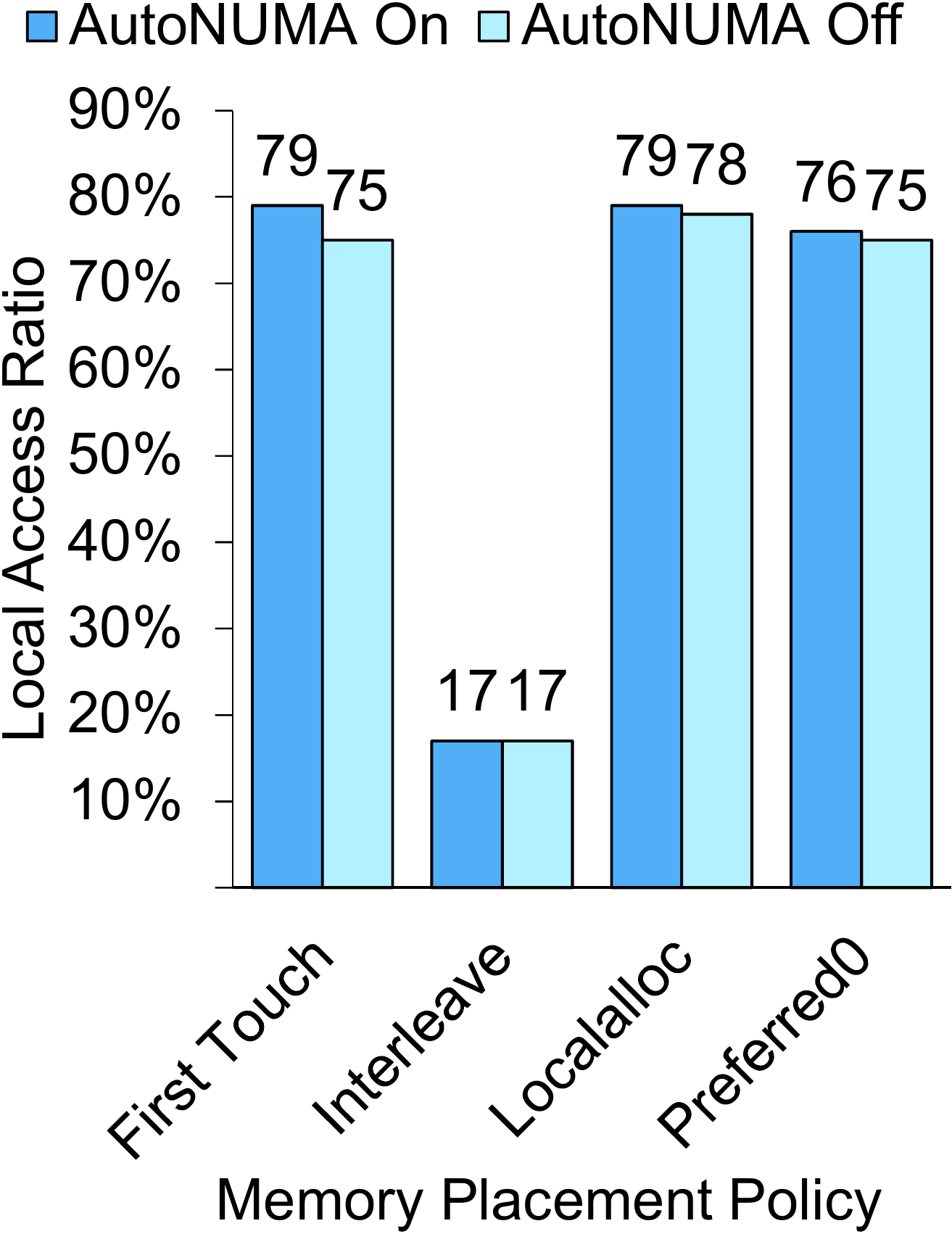}
		\caption{AutoNUMA effect on Local Access Ratio - Machine A}\label{fig:osconfig-autonuma-LAR}
	\end{subfigure}
    \hfill
	\begin{subfigure}[t]{0.205\linewidth}
		\centering
		\includegraphics[width=\linewidth]{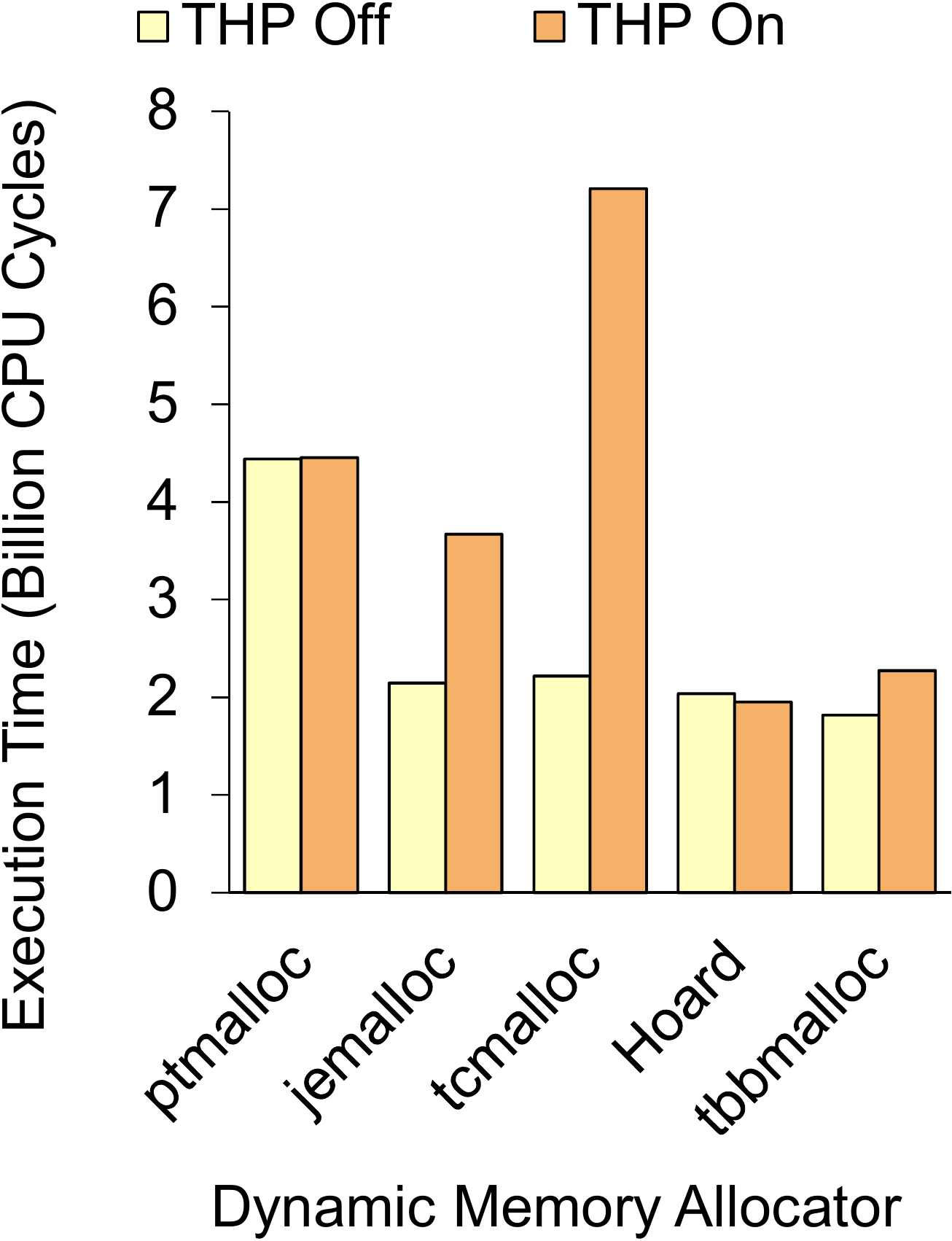}
		\caption{Impact of THP on memory allocators - Machine A}\label{fig:osconfig-THP}
	\end{subfigure}	
	\hfill
	\begin{subfigure}[t]{0.30\linewidth}
		\centering
		\includegraphics[width=\linewidth]{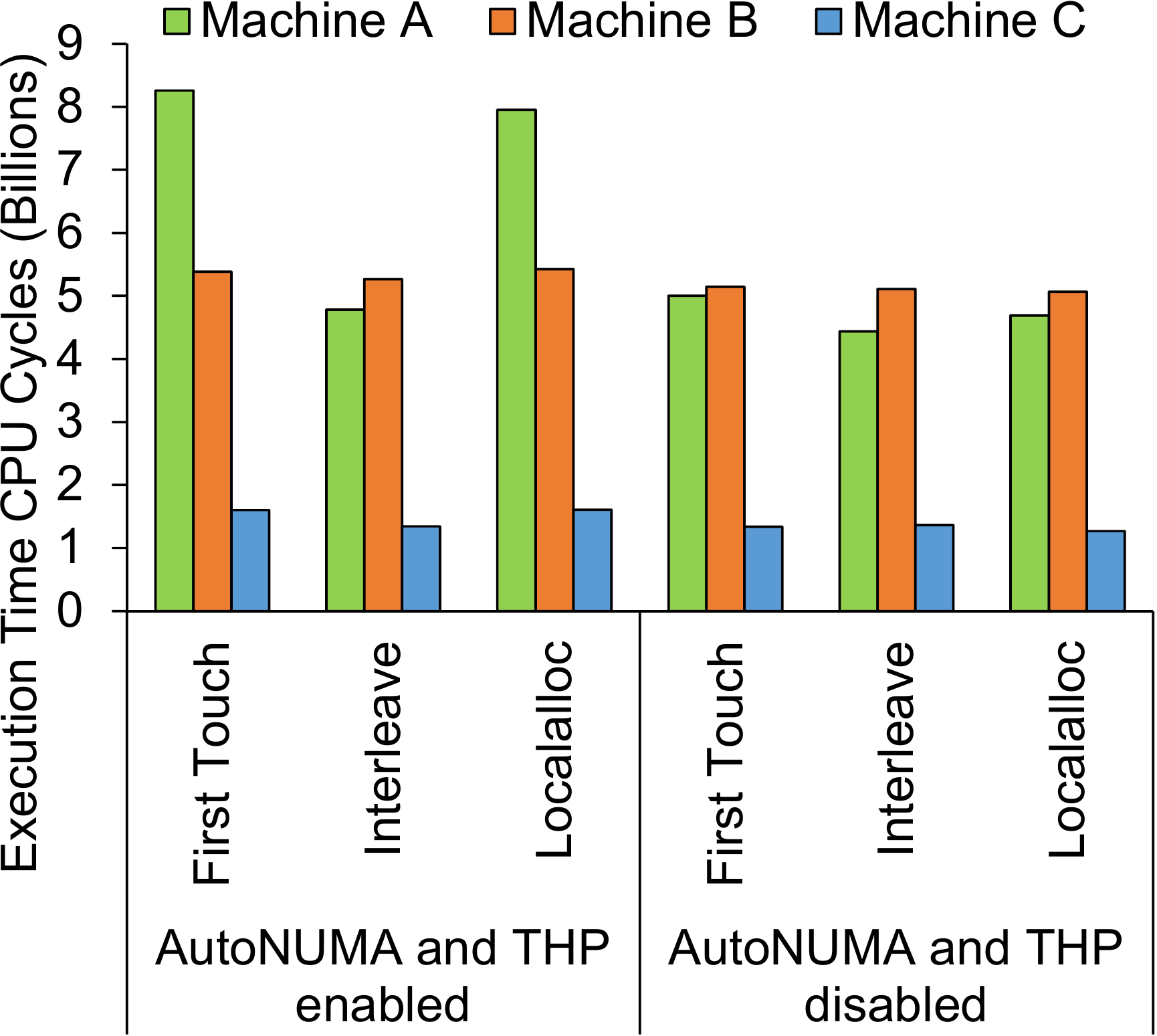}
		\caption{Combined effect of AutoNUMA and THP on different memory placement policies - variable machine}\label{fig:mempolicy-archcomp}
	\end{subfigure}	
	\caption{Impact of operating system configuration (AutoNUMA and THP) on memory placement policies and memory allocators - Holistic Aggregation Workload (W1)}
	\label{fig:osconfig}
	\vspace{-7pt} 
\end{figure*}

\subsection{Datasets and Implementation Details}
\label{subsec:impl}
In this section, we outline the datasets and codebases used for the experiments. We use well-known synthetic datasets outlined in prior work as the basis for all of our experiments ~\cite{cieslewicz2007adaptive,blanas2011design,council2017tpc}.
Unless otherwise noted, all workloads operate on datasets that are stored in memory resident data structures, and any impact from disk I/O is not measured in our results. 

The aggregation workloads (W1 and W2) evaluate a typical hash-based aggregation query, based on a state-of-the-art concurrent hash table ~\cite{li2014algorithmic}, which is implemented as a shared global hash table~\cite{memarzia2019six}. The datasets used for the aggregation workloads are based on three different data distributions: \textit{Moving Cluster}, \textit{Sequential}, and \textit{Zipfian}. In the \textit{Moving Cluster} dataset, the keys are chosen from a window that gradually slides. The \textit{Moving Cluster} dataset provides a gradual shift in data locality that is similar to workloads encountered in streaming or spatial applications. In the \textit{sequential} dataset, we generate a series of segments that contain multiple number sequences. The number of segments is equal to the group-by cardinality, and the number of records in each segment is equal to the dataset size divided by the cardinality. This dataset mimics transactional data where the key incrementally increases. In the \textit{Zipfian} dataset, the distribution of the keys is skewed using Zipf's law \cite{powers1998applications}. We first generate a Zipfian sequence with the desired cardinality $c$ and Zipf exponent $e=0.5$. Then we take $n$ random samples from this sequence to build $n$ records. The Zipfian distribution is used to model many big data phenomena, such as word frequency, website traffic, and city population. For all aggregation datasets, the number of records is 100 million, and the group-by cardinality is one million.

The join workloads (W3 and W4) evaluate a typical join query involving two tables. W3 is a non-partitioning hash join, using the code and dataset from~\cite{blanas2011design}. The dataset contains two tables sized at 16 million and 256 million tuples, and is designed to simulate a decision support system. W4 is an index nested loop join, and uses the same dataset as W3. We evaluated several in-memory indexes for this workload, including ART~\cite{leis2016art} which is based on the concept of a Radix tree and is used in the HyPer~\cite{kemper2011hyper} database, MassTree~\cite{mao2012cache} is a key-value store which uses an indexing technique that is a hybrid of B+Tree and trie, and an in-memory Skip List implementation~\cite{pugh1990skip,skiplist2016}. 

We evaluate a TPC-H workload (W5) on the MonetDB~\cite{monetdb2018} (version 11.33.3) and PostgreSQL~\cite{postgres2019} (version 11.4) databases. MonetDB is an open-source columnar store that uses memory mapped files with demand paging and multiple worker threads for its query processing. PostgreSQL is an open-source row store that uses a volcano-style query processing model. We configured PostgreSQL with a 42GB buffer pool. This workload uses version 2.18 of the industry standard TPC-H dataset specifications. The dataset is designed to mimic a decision support system with eight tables, and is paired with a set of queries which answer typical business questions. Our experiment involves running all 22 queries using a dataset scale factor of 20. We then modify the operating system configuration and run all 22 queries again. Finally, we use Query 5 as a basis for our memory allocator experiment, as it provides a good combination of both joins and aggregation.

The experimental parameters are shown in Table~\ref{tab:expar}. Unless otherwise noted, we use the maximum number of threads supported by each machine. In the synthetic workloads (W1-W4), we measure workload execution time using the timer from~\cite{blanas2011design}. In the TPC-H workload (W5), we use each database system's built-in query timing feature.

\subsection{Operating System Configuration Experiments}
\label{subsec:results-anthp}

In this section, we evaluate three key operating system mechanisms that affect NUMA behavior: NUMA Load Balancing (AutoNUMA), Transparent Hugepages (THP), and the system's memory placement policy. To determine if these variables are affected by other experiment parameters, we also examine the impact of hardware architecture, and the interaction between THP and memory allocators. 

\begin{figure*}[t]
	\centering
	\begin{subfigure}[t]{0.205\linewidth}
		\centering
		\includegraphics[width=\linewidth]{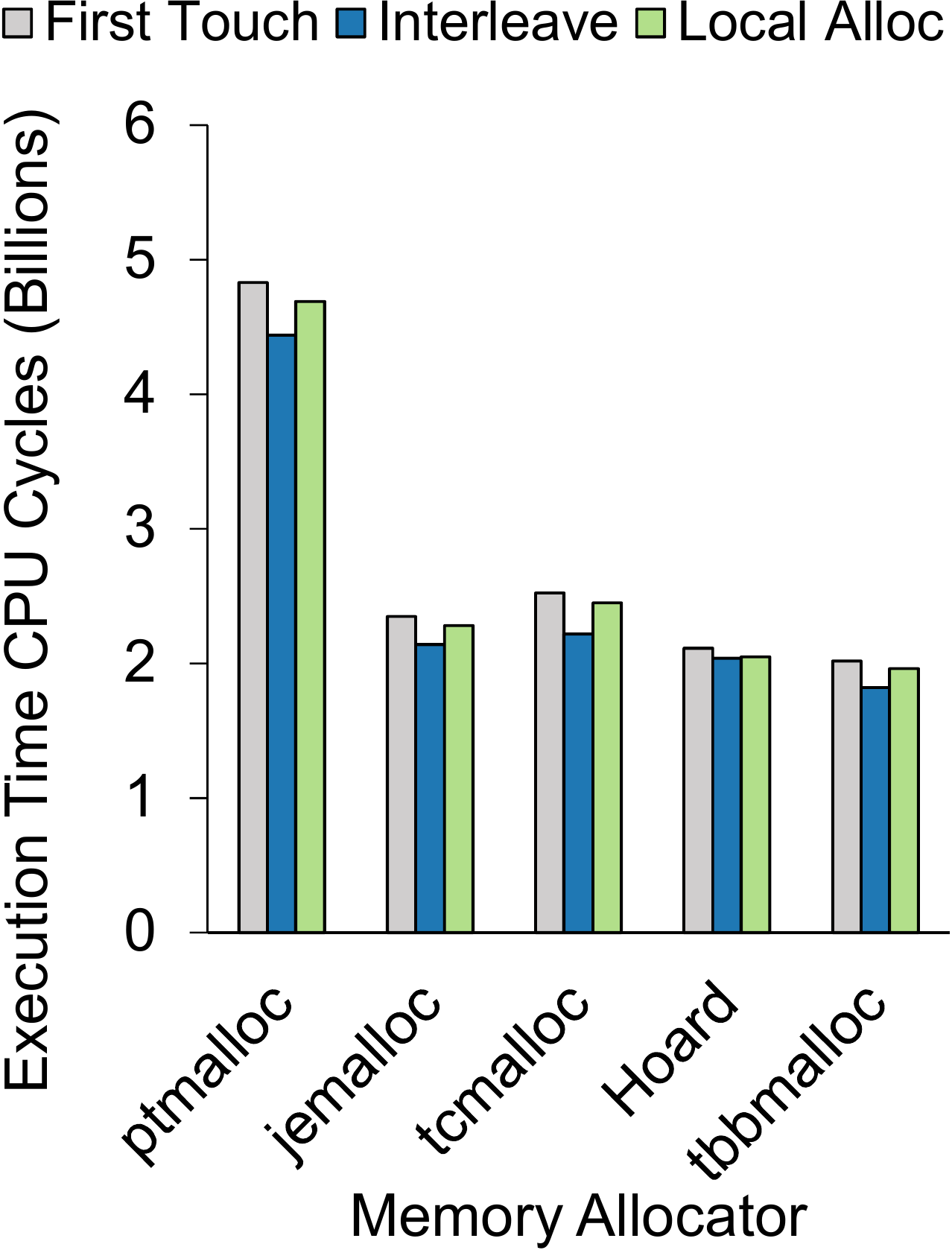}
		\caption{W1 - Machine A}\label{fig:w1-machA}		
	\end{subfigure}
	\hfill
	\begin{subfigure}[t]{0.205\linewidth}
		\centering
		\includegraphics[width=\linewidth]{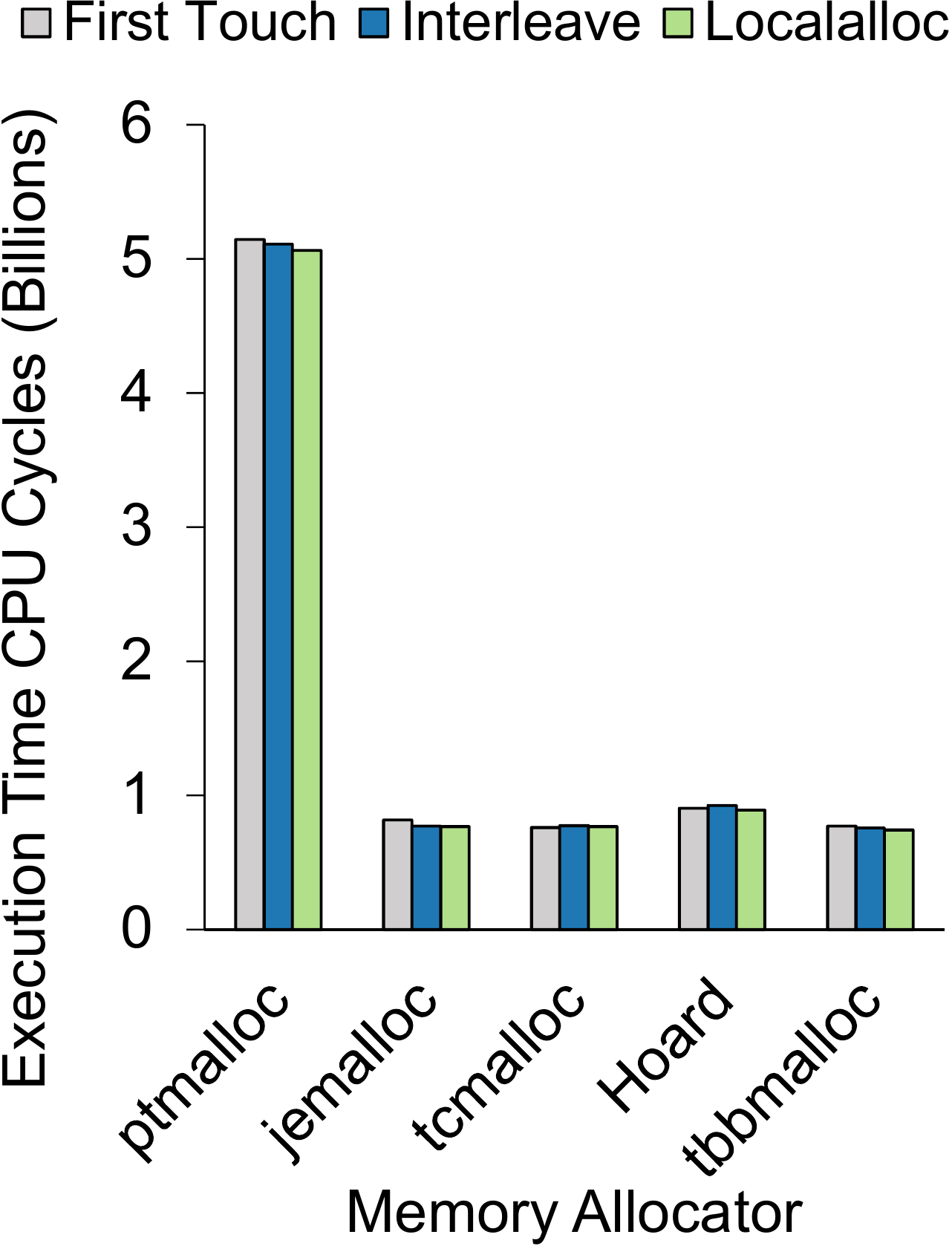}
		\caption{W1 - Machine B}\label{fig:w1-machB}
	\end{subfigure}
	\hfill
    \begin{subfigure}[t]{0.205\linewidth}
		\centering
		\includegraphics[width=\linewidth]{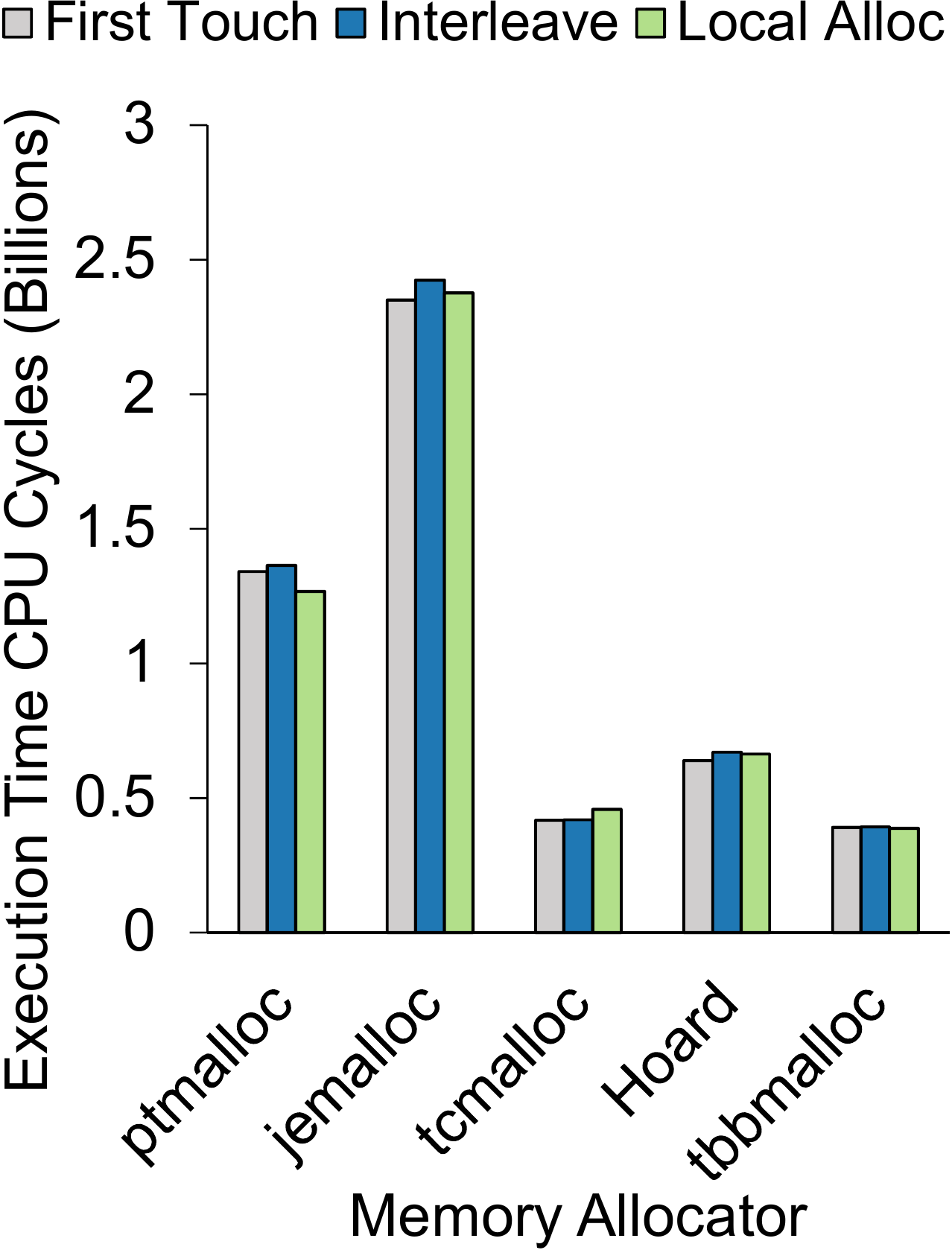}
		\caption{W1 - Machine C}\label{fig:w1-machC}
	\end{subfigure}	
	\hfill
    \begin{subfigure}[t]{0.205\linewidth}
		\centering
		\includegraphics[width=\linewidth]{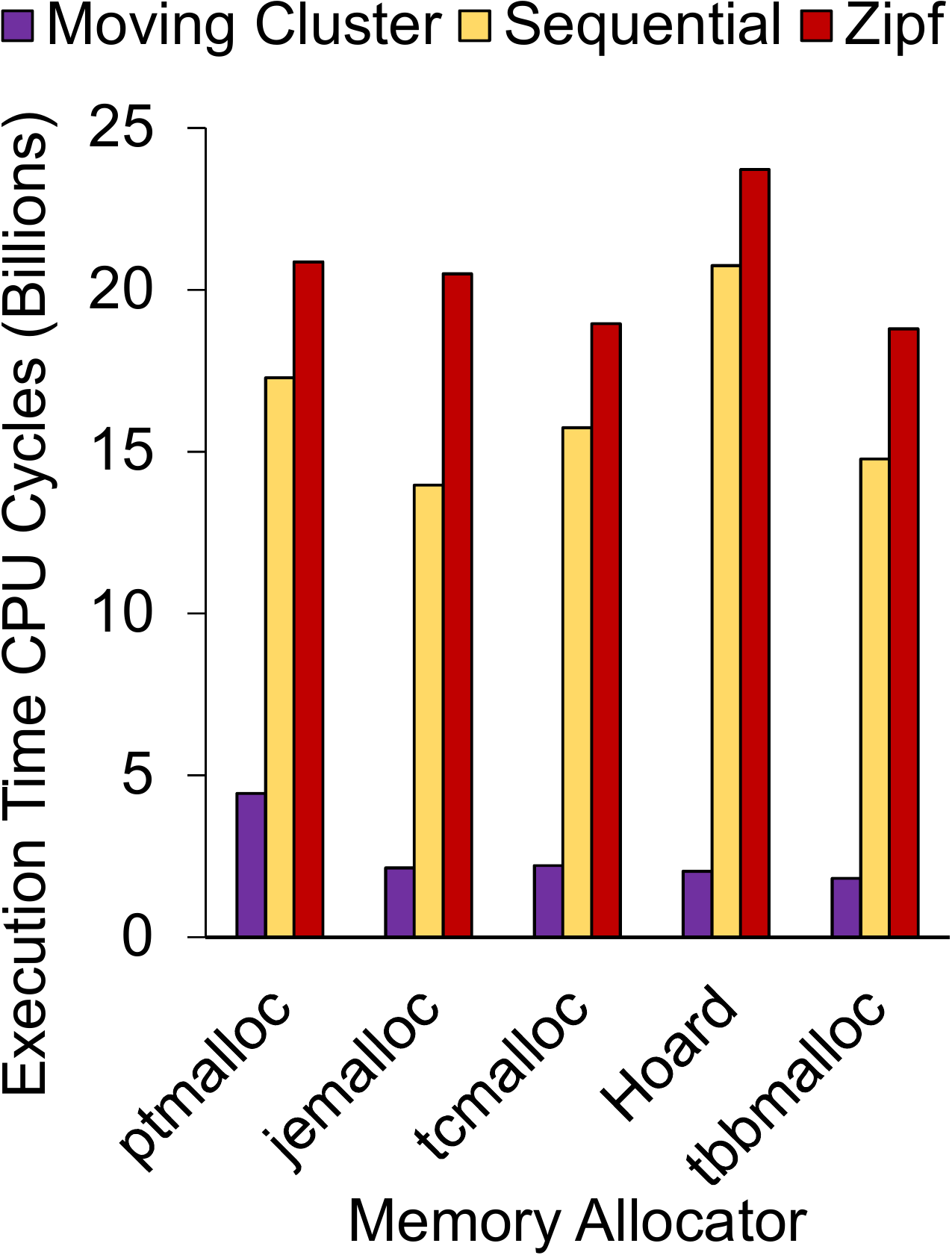}
		\caption{W1 - Machine A - Effect of dataset distribution}\label{fig:vardatset}
	\end{subfigure}
	
    \begin{subfigure}[t]{0.205\linewidth}
		\centering
		\includegraphics[width=\linewidth]{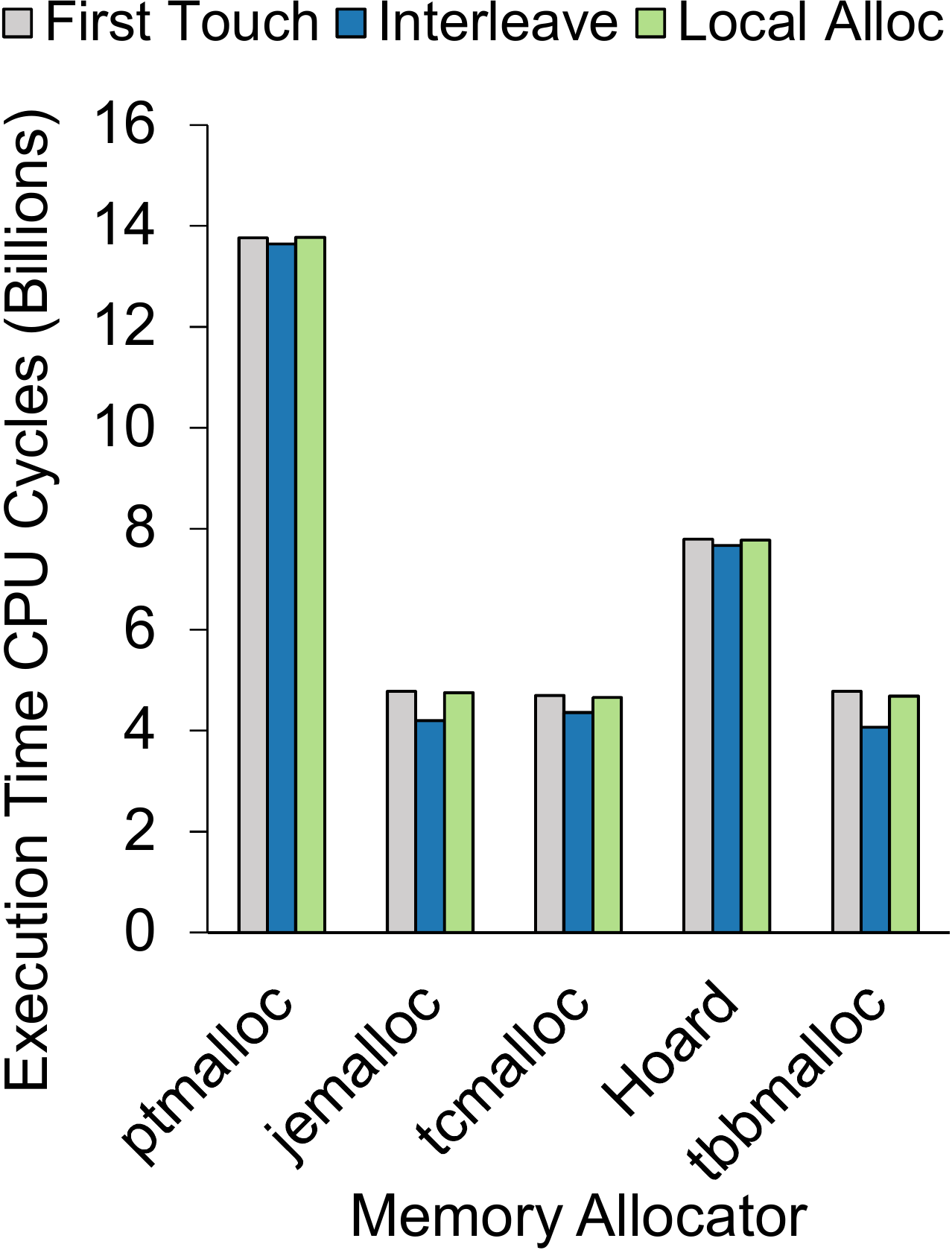}
		\caption{W3 - Machine A}\label{fig:w3-machA}
	\end{subfigure}	
	\hfill
	\begin{subfigure}[t]{0.205\linewidth}
		\centering
		\includegraphics[width=\linewidth]{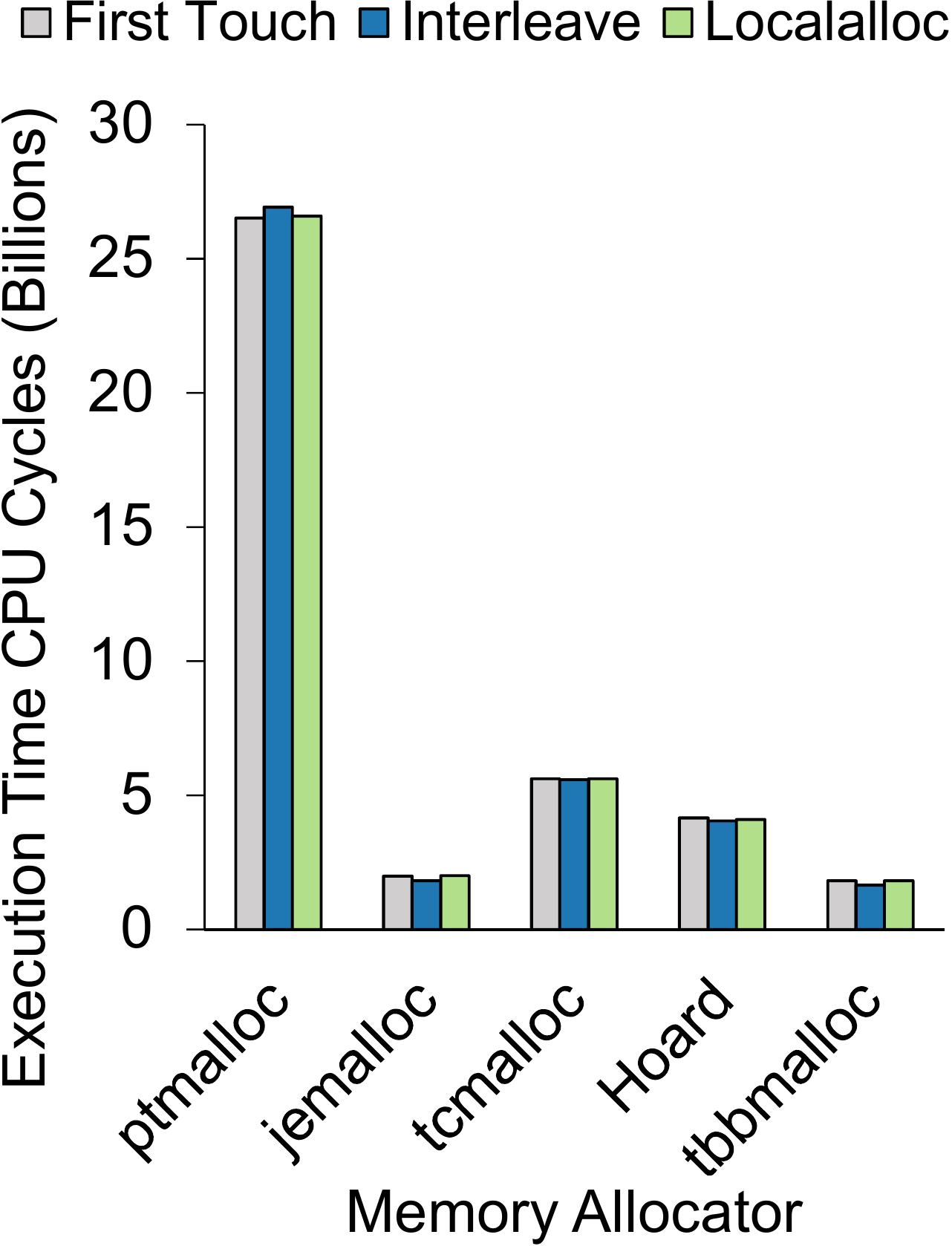}
		\caption{W3 - Machine B}\label{fig:w3-machB}
	\end{subfigure}	
	\hfill
	\begin{subfigure}[t]{0.205\linewidth}
		\centering
		\includegraphics[width=\linewidth]{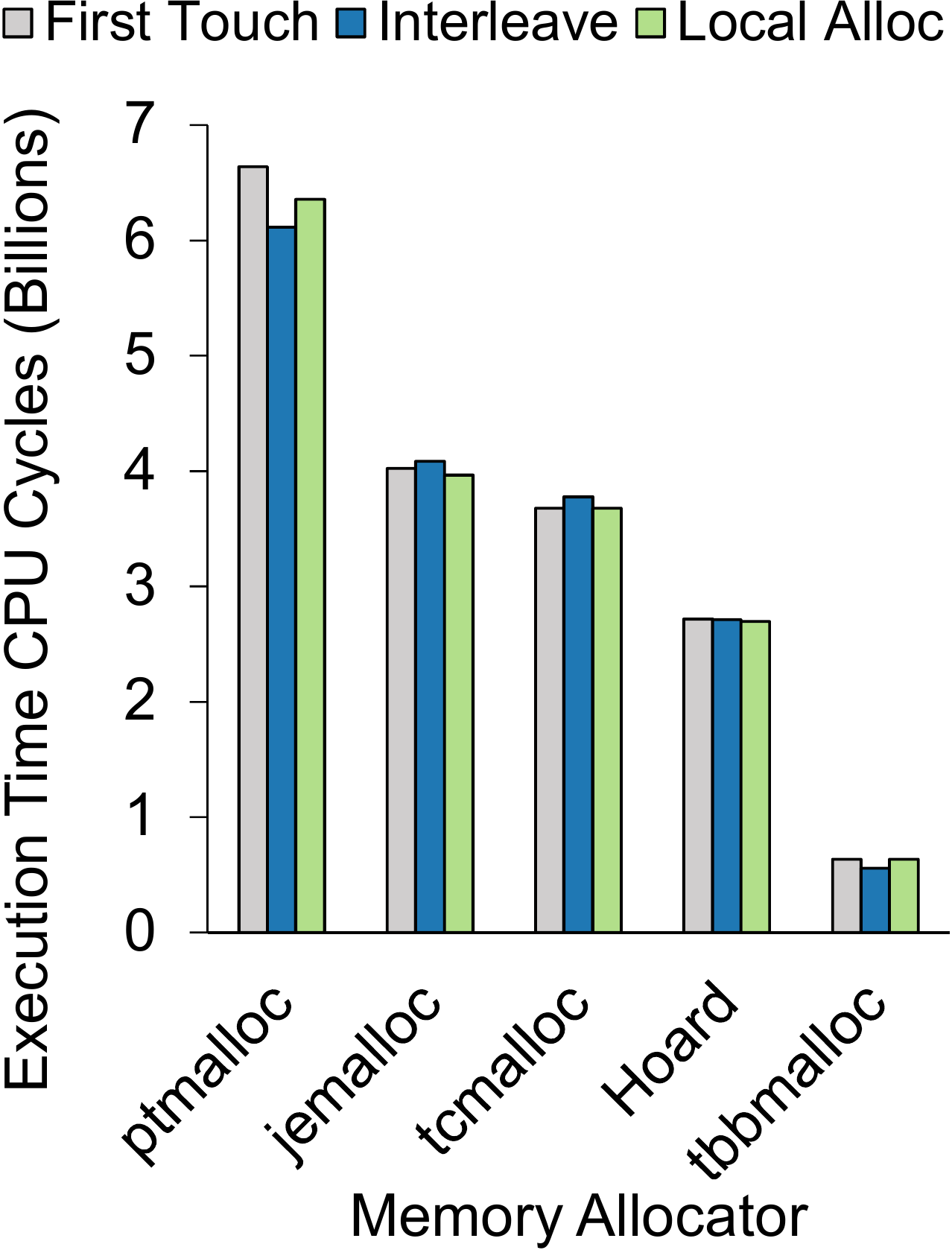}
		\caption{W3 - Machine C}\label{fig:w3-machC}
	\end{subfigure}	
	\hfill
	\begin{subfigure}[t]{0.205\linewidth}
		\centering
		\includegraphics[width=\linewidth]{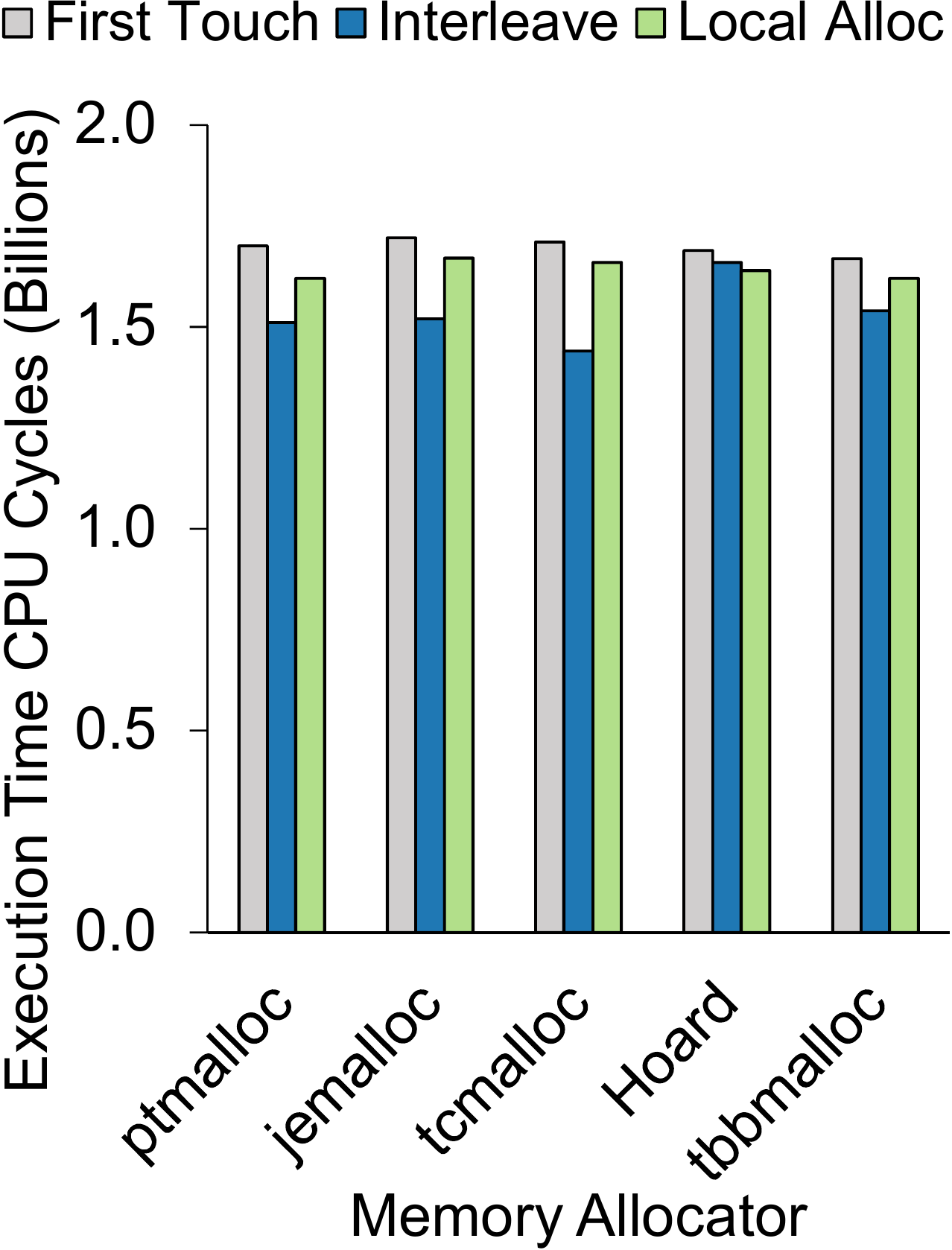}
		\caption{W2 - Machine A}\label{fig:w2-machA-mallocs}
	\end{subfigure}	
	\caption{Comparison of memory allocators - variable memory placement policy}
	\label{fig:memalloc-combined}
\end{figure*}

\subsubsection{AutoNUMA Load Balancing Experiments}
In Figures~\ref{fig:osconfig-autonuma-runtime} and \ref{fig:osconfig-autonuma-LAR}, we evaluate W1 
and toggle the state of AutoNUMA Load Balancing between \textit{On} (the system default) and \textit{Off}. The results in Figure~\ref{fig:osconfig-autonuma-runtime} show that AutoNUMA worsens runtime for the \textit{First Touch}, \textit{Interleave}, and \textit{Localalloc} memory placement policies. Only the \textit{Preferred0} memory placement policy shows an improvement in runtime with AutoNUMA enabled. The \textit{Preferred0} policy tries to allocate memory from NUMA node 0, which is why it benefits the most from AutoNUMA load balancing. These results were obtained using W1 on Machine A, but we observed very similar results on the other workloads and machines. AutoNUMA had a significantly detrimental effect on runtime. The best overall approach is to use memory interleaving and disable AutoNUMA. The \textit{Local Access Ratio (LAR)} shown in Figure~\ref{fig:osconfig-autonuma-LAR} specifies the ratio of memory accesses that were satisfied with local memory \cite{dashti2013traffic} compared to all memory accesses. For example, on our eight node machine, we expect interleaving to result in an average LAR of $100/8=12.5\%$, which is close to our measurement of $17\%$. AutoNUMA's main goal is to improve the LAR by. These results highlight the value of modifying these parameters, as First Touch with load balancing (system default) is $86\%$ slower than Interleave without load balancing.

\subsubsection{Transparent Hugepages Experiments} 
Next we evaluate the effect of the Transparent Hugepages (THP) configuration, which automatically merges groups of 4KB memory pages into 2MB memory pages. As shown in Figure \ref{fig:osconfig-THP}, THP's impact on the workload execution time ranges from detrimental in most cases to a negligible effect in other cases. As THP alters the composition of the operating system's memory pages, support for THP within the memory allocators is the defining factor on whether it is detrimental to performance. \textit{tcmalloc}, \textit{jemalloc}, and \textit{tbbmalloc} are currently not handling THP well. We hope that future versions of these memory allocators will rectify this issue out-of-the-box. Although most Linux distributions enable THP by default, our results indicate that it is generally worthwhile to disable THP for data analytics workloads. 

\begin{figure}[t]
	\centering
	\begin{subfigure}[t]{0.42\linewidth}
		\centering
		\includegraphics[width=\linewidth]{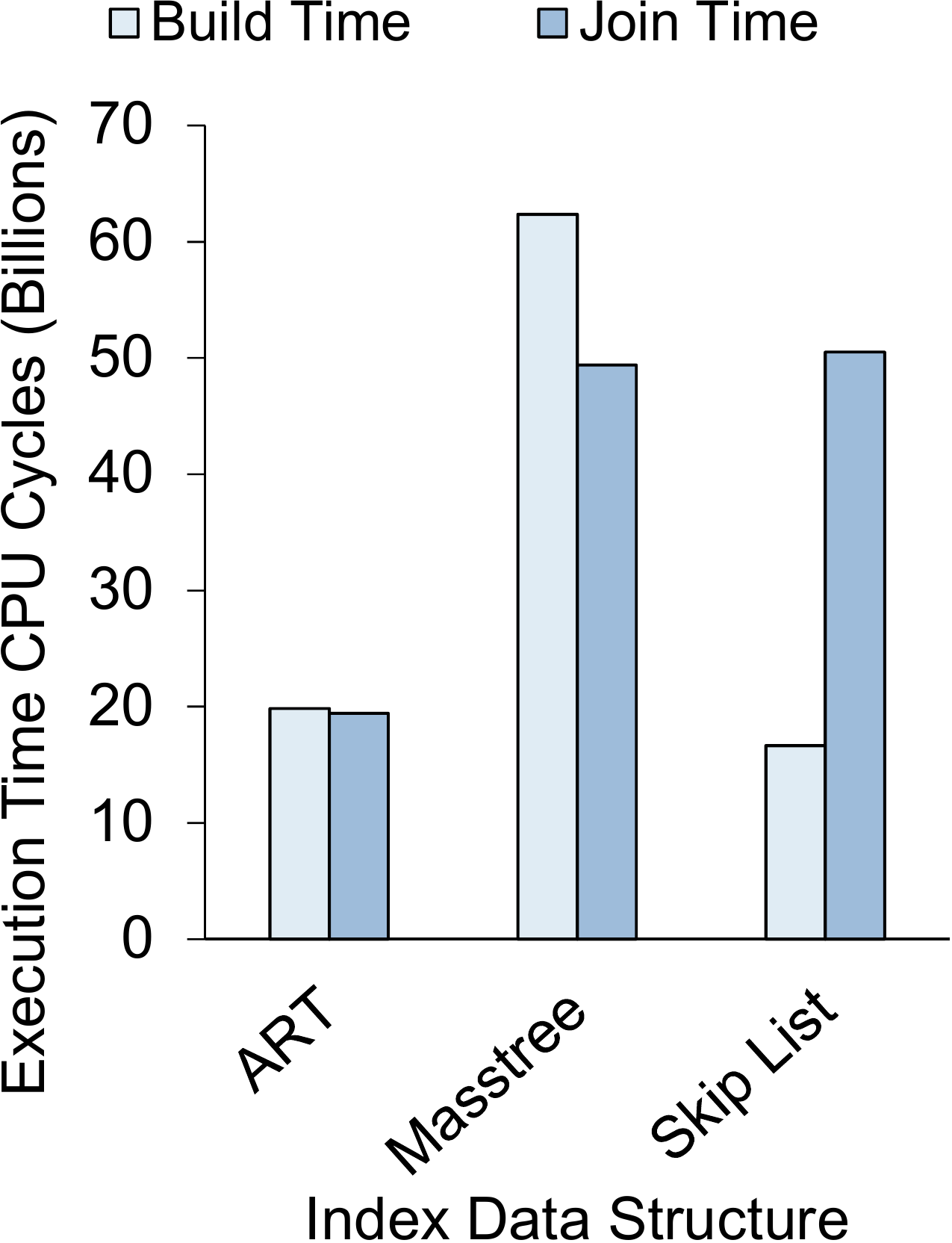}
		\caption{W4 - Index data structure comparison - build and join times}\label{fig:w4-machA-buildandjoin}
	\end{subfigure}	
	\hfill
	\begin{subfigure}[t]{0.42\linewidth}
	\includegraphics[width=\linewidth]{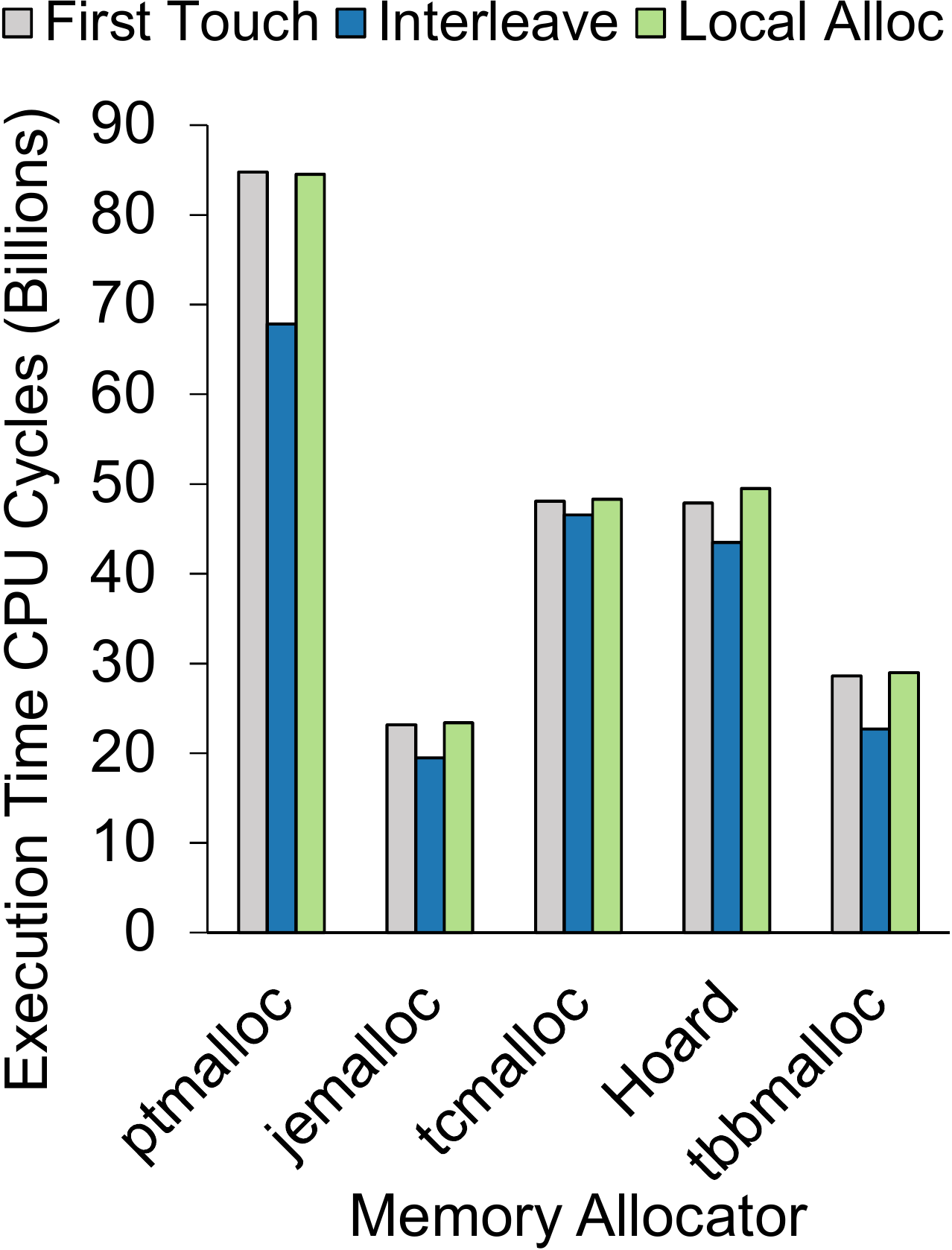}
    \caption{W4 with ART Index - Impact of Memory Allocators}
    \label{fig:w4-machA-mallocs}
    \end{subfigure}
	\caption{Index nested loop join experiments - Machine A}
	\label{fig:indexsel}	
	\vspace{-10pt} 
\end{figure}

\subsubsection{Hardware Architecture Experiments}
\label{subsec:results-hardarch}
Here we show how the performance of data analytics applications running on different machines with different hardware architectures is affected by the memory placement strategies. For all machines, the default configuration uses the \textit{First Touch} memory placement, and both AutoNUMA and THP are enabled. The results depicted in Figure~\ref{fig:mempolicy-archcomp} show that Machine A is slower than Machine B when both machines are using the default configuration. However, using the \textit{Interleave} memory placement policy, and disabling the operating system switches allows Machine A to outperform Machine B by up to 15\%. Machine A shows the most significant improvement from operating system and memory placement policy changes, and the workload runtime is reduced by up to 46\%. The runtime for Machine C is reduced by up to 21\%. The performance improvement on Machine B is around 7\%, which is fairly modest compared to the other machines. Although Machines B and C have a similar inter-socket topology, the relative local and remote memory access latencies are much closer in Machine B (see Table~\ref{tab:machspec}). This, along with other hardware differences, plays a significant role in the benefit gained from altering the memory placement policy. Henceforth, we run our experiments with AutoNUMA and THP disabled, unless otherwise noted.

\subsection{Memory Allocator Experiments}
\label{subsec:results-malloc}

In Section~\ref{subsec:microbench}, we used a memory allocator microbenchmark to show that there are significant differences in both multi-threaded scalability and memory consumption overhead. In this section, we explore the performance impact of overriding the system default memory allocator, using four in-memory data analytics workloads. These experiments aim to reveal the relationship between workload, hardware architecture, and memory allocator. 

\subsubsection{Hashtable-based Experimental Workloads}

In Figure \ref{fig:memalloc-combined}, we show our results for the holistic aggregation (W1), distributive aggregation (W2), and hash join (W3) workloads, running on each of our three machines. In addition to the memory allocators, we vary the memory placement policies for each workload. The results show significant runtime reductions on all three machines, particularly when using \textit{tbbmalloc} in conjunction with the \textit{Interleave} memory placement policy. The holistic aggregation workload (W1) shown in Figure \ref{fig:w1-machA} to \ref{fig:w1-machC} extensively uses memory allocation during its runtime to store the tuples for each group and calculate their aggregate value. Utilizing \textit{tbbmalloc} reduced the runtime of W1 by up to $62\%$ on machine A, $83\%$ on machine B, and $72\%$ on machine C, compared to the default allocator (\textit{ptmalloc}). The results for the join query (W3) depicted in Figures \ref{fig:w3-machA} to \ref{fig:w3-machC} also show significant improvements, with \textit{tbbmalloc} reducing workload execution time by $70\%$ on machine A, $94\%$ on machine B, and $92\%$ on machine C. The distributive aggregation query (W2) shown in Figure~\ref{fig:w2-machA-mallocs} does not gain much of a benefit, as it calculates a running count using a hash table, and is therefore comparatively light on memory allocation.

\subsubsection{Impact of Dataset Distribution}
The performance of query workloads and memory allocators can be sensitive to the access patterns induced by the dataset distribution. The datasets are the same size and their key differentiating factor is the way their records are distributed (see Section~\ref{subsec:impl} for more information). In our previous figures, we used the \textit{Heavy Hitter} dataset as the default dataset for W1. In Figure~\ref{fig:vardatset}, we vary the dataset to see if overriding the default memory allocator is still beneficial. With the exception of the \textit{Hoard} allocator, all of the alternative memory allocators improve W1's runtime on the \textit{Zipf} and \textit{Sequential} datasets. In particular, \textit{jemalloc} and \textit{tbbmalloc} provide the largest benefits.

\subsubsection{Effect on In-memory Indexing}
The index used to accelerate the nested loop join workload (W4) plays a major role in determining its efficiency. Although there are many data structures that could be used for indexing, efficient concurrency is less trivial to implement. In Figure~\ref{fig:w4-machA-buildandjoin}, we evaluate the time to build the index and the time to run the join workload for three in-memory indexes: ART~\cite{leis2016art}, Masstree~\cite{mao2012cache}, and Skip List~\cite{skiplist2016}. Based on the results, we select ART as the index with the best overall performance. In W4, we are interested in the join time, given a pre-built index. We included the build time as an interesting sidenote, since ART performs well in this regard as well. In Figure~\ref{fig:w4-machA-mallocs}, we show the beneficial effect of overriding the memory allocators for W4 when using the ART index. The reduction in runtime is substantial, particularly with the \textit{jemalloc} memory allocator, and further performance gains are obtained from memory interleaving.  

\begin{figure}[t]
    \includegraphics[width=\linewidth]{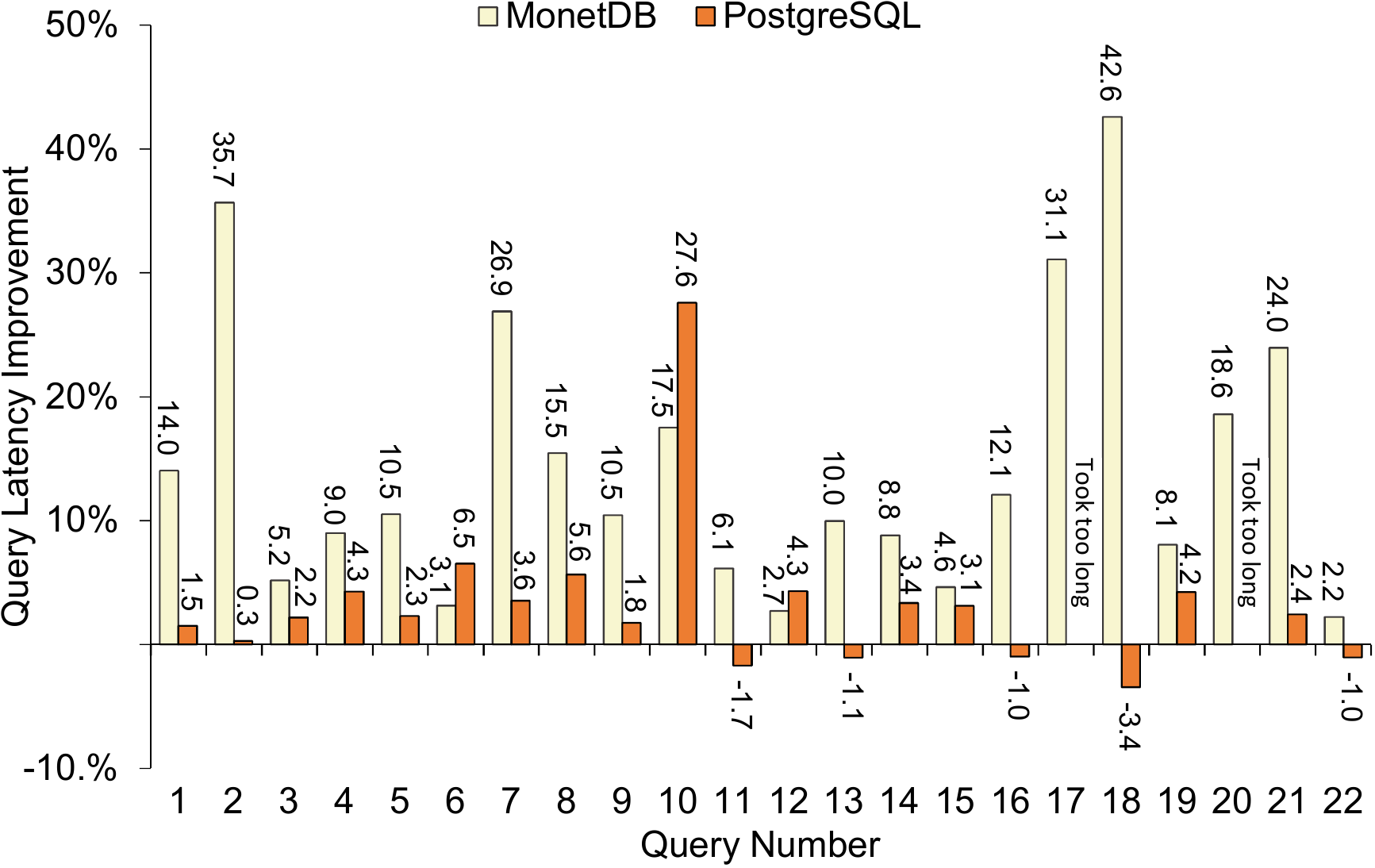}
    \vspace{+0.05cm}
    \caption{Query latency improvement gained from disabling AutoNUMA and THP - all 22 queries - Machine A} 
    \label{fig:monetdb-anthp-22queries}
\end{figure}

\begin{figure}[t]
	\centering
	\begin{subfigure}[t]{0.42\linewidth}
	\includegraphics[width=\linewidth]{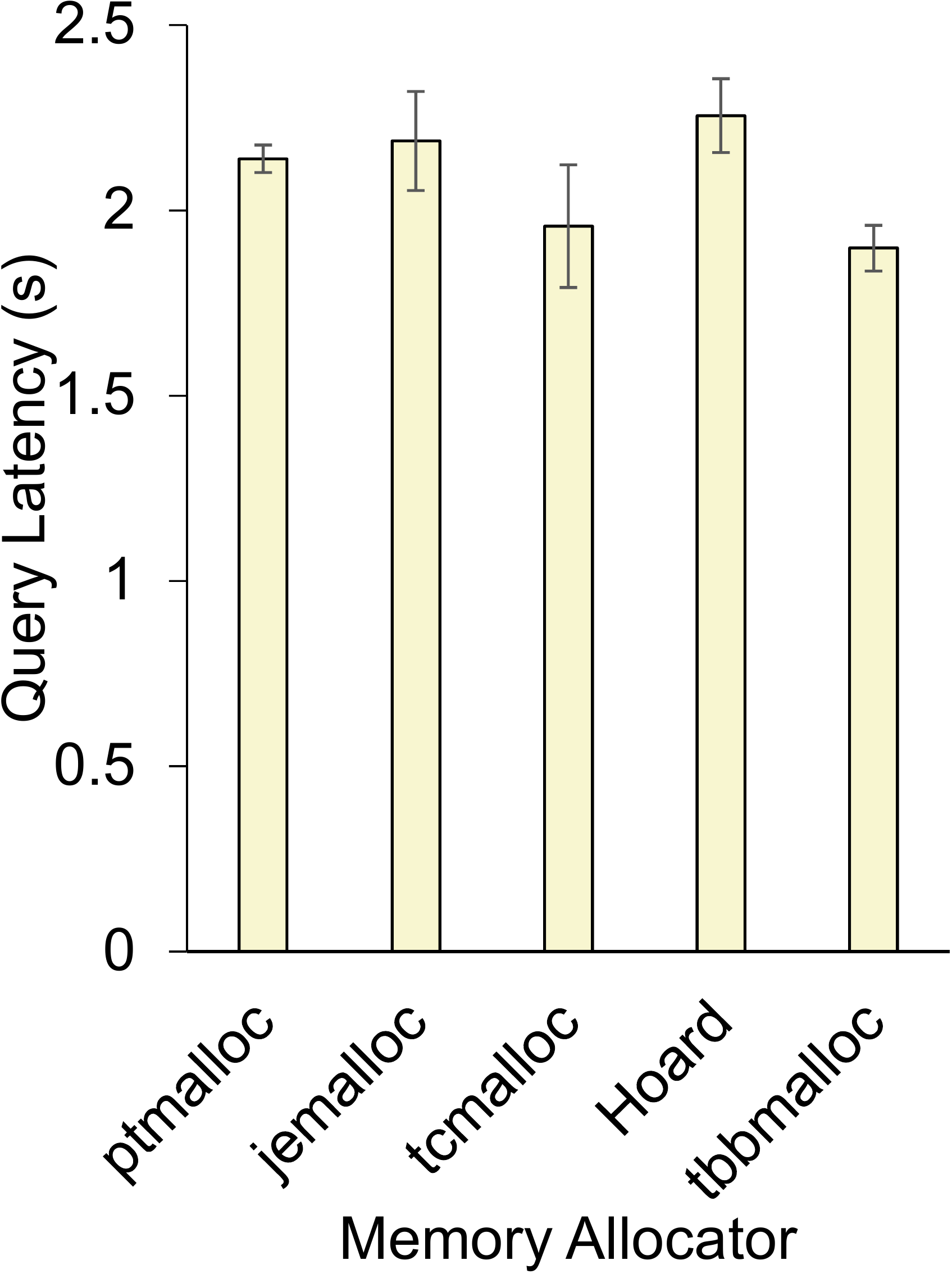}
    \caption{Query 5}
    \label{fig:monetdb-malloc-Q5}
    \end{subfigure}
    \hfill
	\begin{subfigure}[t]{0.42\linewidth}
		\centering
		\includegraphics[width=\linewidth]{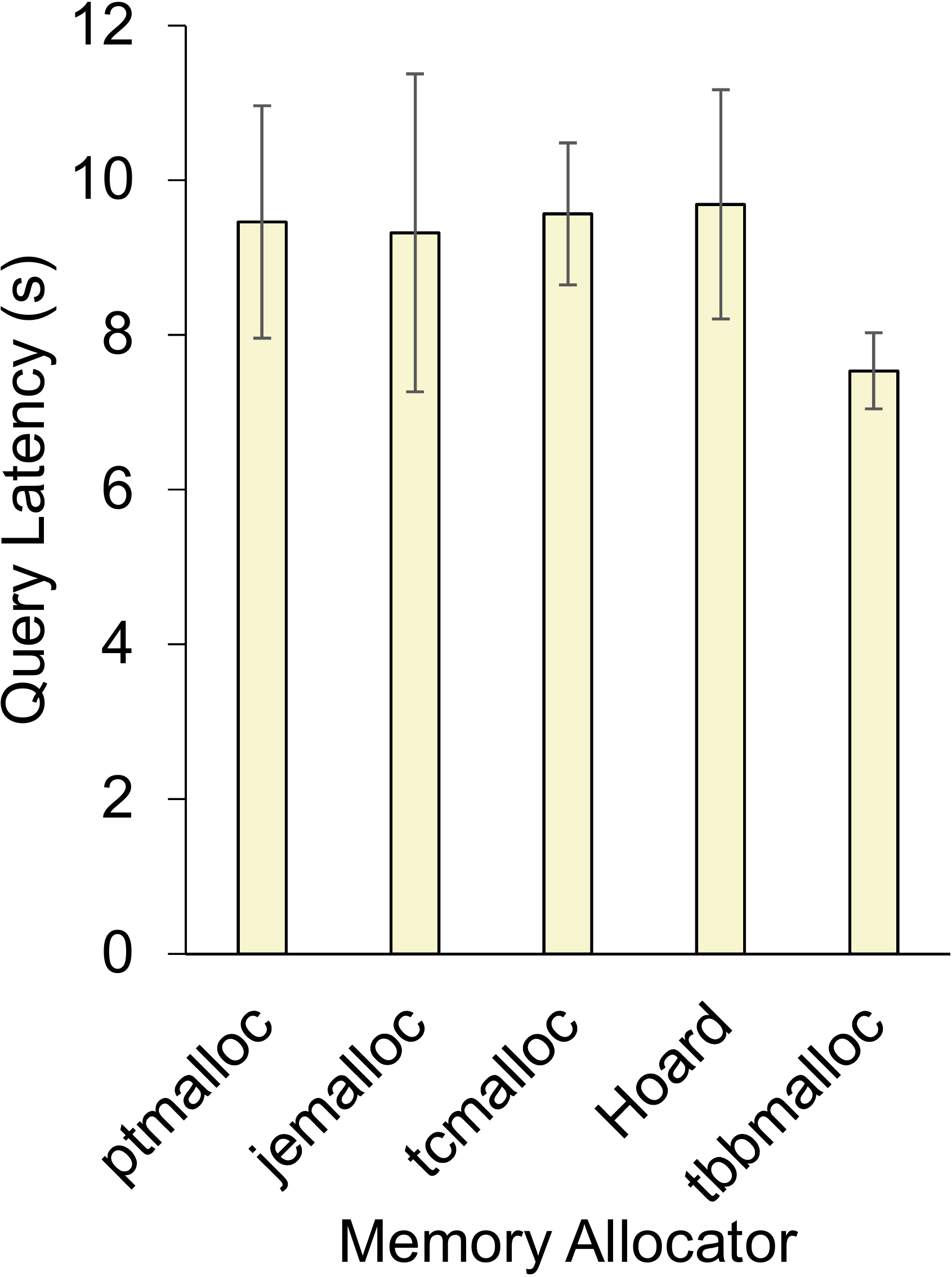}
		\caption{Query 18}\label{fig:monetdb-malloc-Q18}
	\end{subfigure}
	\vspace{+0.1cm}
	\caption{Effect of memory allocator on TPC-H query latency - MonetDB - Machine A}
	\label{fig:monetdb-malloc}	
\end{figure}

\subsection{Database Engine Experiments}
\label{subsec:results-tpch}
In this section, we analyze the the TPC-H workload (W5) on two database systems. Measuring NUMA-related effects on database systems like MonetDB or PostgreSQL is more difficult compared to synthetic workloads, as the database systems are loading data on demand from the disk rather than keeping all the data memory resident. To ensure fair results, we clear the page cache before running the workload and report the average runtime, after disregarding the first (cold) run. In a similar vein to the other experiments, we evaluated the impact of the operating system configuration, memory placement policies, and memory allocators. First we ran all 22 TPC-H queries, and calculated the query latency reduction caused by disabling AutoNUMA and THP, compared to the system default. The results depicted in Figure~\ref{fig:monetdb-anthp-22queries} show that MonetDB's query latencies improved between 2\% and 43\%, with an average improvement of 14.5\%. The results for PostgreSQL are less impressive, with an average improvement of 3\% and five queries taking longer to complete. We believe this is due to PostGreSQL's rigid multi-process query processing approach. Next we evaluate the effect of memory allocator overriding on MonetDB. To do so, we selected queries 5 and  18 due to their usage of both joins and aggregation. The results shown in Figure~\ref{fig:monetdb-malloc-Q5} indicate that \textit{tbbmalloc} can provide an average query latency reduction of up to 12\% for Query 5, and 20\% for Query 18, compared to \textit{ptmalloc}. 

\subsection{Summary}
\label{sec:summary}

The strategies outlined in this paper, when carefully applied, can significantly speed up data analytics workloads without the need for modifying the application source code. The effectiveness and applicability of these strategies to a workload depend on several factors. Starting with the operating system configuration, we showed that the default settings for AutoNUMA and THP can have a significant detrimental effect on performance. AutoNUMA's overhead has proven to be too costly for multi-threaded data analytics workloads. THP provides no benefit to these workloads because they rely on random rather than contiguous memory access patterns. Furthermore, some memory allocators do not support THP, potentially resulting in dramatic performance drops. Although root access is required to access the AutoNUMA setting, we observed that that the \textit{Interleave} memory policy (which can be used by a regular user) can largely nullify AutoNUMA's negative impact. We noted in our evaluation that the effects of the memory placement policies are less pronounced when AutoNUMA is disabled, with Machine A obtaining the most benefit from interleaved memory placement. Different dynamic memory allocators have targeted different use cases and systems, and our microbenchmark showed significant differences in terms of scalability and efficiency. In our evaluation, we demonstrated that these differences translate into real gains in data analytics workloads. Deciding whether to use an alternative memory allocator depends on the answer to the following question: does my workload frequently involve multiple threads concurrently allocating memory? If the answer is yes, then memory allocators are an avenue worth exploring for the application. We believe the combination all of these findings can provide guidance to developers and practitioners.
\section{Related Work} 
\label{sec:related}

The rising demand for high performance parallel computing has motivated many works on leveraging NUMA architectures. We now explore some of works in this context that are relevant to query processing and data analytics. 

In~\cite{kiefer2013experimental}, Kiefer et al. evaluated the performance impact of NUMA effects on multiple independent instances of the MySQL database system. They explored several experiment parameters, including different thread and memory assignment strategies. The authors noted that cache locality and remote access can be harmful to query performance. 

In recent years, there have been several projects to design load balancing approaches that can automatically improve NUMA system performance in an application-agnostic manner. These approaches generally focus on improving performance by altering the process and/or memory placement. Many researchers have pursued the philosophy that tuning applications for better NUMA performance should be possible without modifying the code. There are two subcategories of this research in this area: (a) load balancing daemons that co-exist with existing operating systems and provide guided thread scheduling and memory management, b) custom operating systems designed for multicore architectures and/or database system. Some examples of the first category include Dino \cite{blagodurov2010case}, Carrefour \cite{dashti2013traffic}, AsymSched \cite{lepers2015thread}, Numad \cite{redhat2018}, and AutoNUMA \cite{redhat2018}. These schedulers have been shown to improve performance in some cases, particularly on systems running multiple independent processes. However, other researchers have claimed that these schedulers do not provide much benefit for multi-threaded in-memory query processing applications \cite{psaroudakis2015scaling, schuh2016experimental}.

\balance

The effects of operating system behavior on data processing workloads have led some researchers to pursue the creation of custom-tailored operating systems for database applications~\cite{giceva2019os,giceva2016customized,giceva2014deployment,giceva2012towards}. For example, Giceva et al.~\cite{giceva2016customized} developed a light-weight kernel for the Barrelfish~\cite{baumann2009multikernel} operating system. This modified operating system is designed to provide the minimal requirements to run a database system. The authors propose the option for task-based scheduling. Unlike threads, tasks are given dedicated access to a processor, and they will not be interrupted or preempted by the operating system. The authors demonstrated runtime improvements for three graph processing queries, when run inside a noisy multi-programming environment. 

Another approach is to integrate NUMA-oriented features into data structures, but leave the application of these features up to the developer. This solution is not automatic, but can make it easier for developers to adapt their application to different target systems. Psaroudakis et al.~\cite{psaroudakis2018analytics} propose a smart array that has several NUMA-oriented features baked into the data structure. The smart array can be configured to replicate itself across multiple NUMA nodes, interleave its data, or relocate to a particular node.

Some works have explored application-oriented approaches that fine-tune query processing algorithms to the hardware. Wang et al. \cite{wang2015numa} proposed an aggregation algorithm for NUMA systems, based on radix partitioning. Our concurrent aggregation implementation is similar to the shared aggregation (SA) approach described in their work, which allows parallel instances to share a global hash table, and leverages task stealing for improved load balancing. Hash joins typically follow similar a pattern to hash-based aggregation, and are affected by many of the challenges that affect aggregation on NUMA systems. Leis et al. \cite{leis2014morsel} presented a NUMA-aware parallel scheduling algorithm for hash joins. Their approach uses dynamic task stealing in order to deal with dataset skew. Schuh et al. \cite{schuh2016experimental} conducted an in-depth analysis of thirteen main memory join algorithms on a NUMA system. The authors conducted a theoretical and empirical analysis of the algorithms. They concluded that partitioning is required to achieve good performance, unless the probe relation is highly skewed. However, the authors only considered skew on the probe relation, overlooking the possibility of skew on the build relation. Wang et al. \cite{wang2015numa} proposed load balancing aggregation algorithms for NUMA systems. Similarly to \cite{leis2014morsel}, they use a single global hash table, and they only allow inter-socket task stealing when all tasks assigned to a particular socket have been completed. Researchers have also investigated data partitioning in the context of NUMA-aware in-memory storage~\cite{kissinger2014eris, Porobic14ATraPos}. Psaroudakis et al.~\cite{Psaroudakis2016AdaptiveNUMA} developed techniques for adaptive data placement and work-stealing to fix imbalance in resource utilization. Our work is orthogonal to these approaches and they can benefit from using the application-agnostic strategies that we suggest.
\section{Conclusion}
\label{sec:conc}

In this work, we have provided empirical evidence and analysis to support the importance of using application-agnostic strategies to speedup data analytics workloads on NUMA machines. Our experiments on five analytics workloads have shown that it is possible to obtain significant speedups by utilizing these strategies. We also observed that current operating system default configurations are generally sub-optimal for in-memory data analytics. Our results, surprisingly, indicate that operating system features, such as AutoNUMA and Transparent Hugepages, should be disabled for data analytics workloads, regardless of the hardware generation. Furthermore, a lack of thread affinitization was shown to produce inconsistent results and severe performance penalties, indicating possible shortcomings in the default operating system behavior. We found that memory page interleaving generally provided the fastest runtimes out of all the memory placement strategies. We used a microbenchmark to show that memory allocator scalability on NUMA systems should be considered, and that this under-appreciated topic is ripe for investigation. We then obtained large speedups for our data analytics workloads by overriding the default dynamic memory allocator with alternatives, such as \textit{tbbmalloc} and \textit{jemalloc}. 

As our approach does not target a specific NUMA topology, we have shown that our findings can be applied to systems with different architectures. As hardware architectures continue to advance towards greater parallelism and greater levels of memory access partitioning, we hope this work can help practitioners address similar issues. 

As our future work, we plan to delve deeper into memory allocator fine-tuning, by tweaking the allocators themselves. We would also like to study the new challenges and opportunities presented by emerging hardware, such as multi-socket multi-chip-module systems, which are going to create an even larger hierarchy of memory access latencies. Lastly, we are considering developing a tool that can automatically set good system defaults (memory allocator, placement policy, and schedulers) without requiring much tuning.
\section{Acknowledgements}
\label{sec:ack}

We would like to thank Kenneth Kent and Aaron Graham from IBM CASA and Serguei Vassiliev and Kaizaad Bilimorya from Compute Canada, for providing access to Machine B and Machine C respectively.

\bibliographystyle{abbrv}
\bibliography{main.bib}

\begin{thebibliography}{10}

\bibitem{asanovic2006landscape}
K.~Asanovic et~al.
\newblock The landscape of parallel computing research: A view from berkeley.
\newblock Technical report, Technical Report UCB/EECS-2006-183, University of
  California, Berkeley, 2006.

\bibitem{balkesen2013main}
C.~Balkesen et~al.
\newblock {Main-memory hash joins on multi-core CPUs: Tuning to the underlying
  hardware}.
\newblock In {\em ICDE}, 2013.

\bibitem{Baumann:2009:TheMultikernel}
A.~Baumann, P.~Barham, P.-E. Dagand, T.~Harris, R.~Isaacs, S.~Peter, T.~Roscoe,
  A.~Sch\"{u}pbach, and A.~Singhania.
\newblock The multikernel: A new os architecture for scalable multicore
  systems.
\newblock In {\em ACM SIGOPS Symposium on Operating Systems Principles (SOSP)},
  SOSP '09, pages 29--44, 2009.

\bibitem{baumann2009multikernel}
A.~Baumann, P.~Barham, P.-E. Dagand, T.~Harris, R.~Isaacs, S.~Peter, T.~Roscoe,
  A.~Sch{\"u}pbach, and A.~Singhania.
\newblock {The multikernel: a new OS architecture for scalable multicore
  systems}.
\newblock In {\em ACM SIGOPS Symposium on Operating Systems Principles (SOSP)},
  pages 29--44. ACM, 2009.

\bibitem{berger2000hoard}
E.~D. Berger et~al.
\newblock Hoard: A scalable memory allocator for multithreaded applications.
\newblock {\em ACM SIGARCH}, 28(5), 2000.

\bibitem{DBLP:conf/pldi/BergerZM01}
E.~D. Berger, B.~G. Zorn, and K.~S. McKinley.
\newblock Composing high-performance memory allocators.
\newblock In Burke and Soffa \cite{DBLP:conf/pldi/2001}, pages 114--124.

\bibitem{blagodurov2010case}
S.~Blagodurov et~al.
\newblock {A case for NUMA-aware contention management on multicore systems}.
\newblock In {\em PACT}, 2010.

\bibitem{blanas2011design}
S.~Blanas et~al.
\newblock {Design and evaluation of main memory hash join algorithms for
  multi-core CPUs}.
\newblock In {\em SIGMOD}, 2011.

\bibitem{Boyd-Wickizer:2008:Corey}
S.~Boyd-Wickizer, H.~Chen, R.~Chen, Y.~Mao, F.~Kaashoek, R.~Morris,
  A.~Pesterev, L.~Stein, M.~Wu, Y.~Dai, Y.~Zhang, and Z.~Zhang.
\newblock Corey: An operating system for many cores.
\newblock In {\em USENIX Conference on Operating Systems Design and
  Implementation (OSDI)}, OSDI'08, pages 43--57, 2008.

\bibitem{DBLP:conf/pldi/2001}
M.~Burke and M.~L. Soffa, editors.
\newblock {\em {ACM} {SIGPLAN} Conference on Programming Language Design and
  Implementation (PLDI)}. {ACM}, 2001.

\bibitem{chen2017contention}
Q.~Chen and M.~Guo.
\newblock Contention and locality-aware work-stealing for iterative
  applications in multi-socket computers.
\newblock {\em IEEE Transactions on Computers}, 67(6):784--798, 2017.

\bibitem{cieslewicz2007adaptive}
J.~Cieslewicz and K.~A. Ross.
\newblock Adaptive aggregation on chip multiprocessors.
\newblock In {\em VLDBJ}, 2007.

\bibitem{corbet2012autonuma}
J.~Corbet.
\newblock {AutoNUMA: the other approach to NUMA scheduling}.
\newblock {\em LWN. net}, 2012.

\bibitem{council2017tpc}
T.~P.~P. Council.
\newblock {TPC-H benchmark specification 2.17.3}.
\newblock {\em Published at tpc.org/tpch}, 2017.

\bibitem{dashti2013traffic}
M.~Dashti et~al.
\newblock {Traffic management: a holistic approach to memory placement on NUMA
  systems}.
\newblock {\em SIGPLAN Notices}, 48(4), 2013.

\bibitem{dice2015influence}
D.~Dice, T.~Harris, A.~Kogan, and Y.~Lev.
\newblock {The influence of malloc placement on TSX hardware transactional
  memory}.
\newblock {\em arXiv preprint arXiv:1504.04640}, 2015.

\bibitem{diener2017affinity}
M.~Diener et~al.
\newblock Affinity-based thread and data mapping in shared memory systems.
\newblock {\em CSUR}, 49(4), 2017.

\bibitem{evans2006scalable}
J.~Evans.
\newblock {A scalable concurrent malloc (3) implementation for FreeBSD}.
\newblock In {\em BSDCan}, 2006.

\bibitem{jason2017}
J.~Evans.
\newblock Jemalloc wiki.
\newblock \url{github.com/jemalloc/jemalloc/wiki/Background}, 2017.

\bibitem{funston2018placement}
J.~Funston, M.~Lorrillere, A.~Fedorova, B.~Lepers, D.~Vengerov, J.-P. Lozi, and
  V.~Quema.
\newblock Placement of virtual containers on {NUMA} systems: A practical and
  comprehensive model.
\newblock In {\em {USENIX} Annual Technical Conference ({USENIX} {ATC} 18)},
  pages 281--294, Boston, MA, 2018. {USENIX} Association.

\bibitem{gaud2015challenges}
F.~Gaud, B.~Lepers, J.~Funston, M.~Dashti, A.~Fedorova, V.~Qu{\'e}ma,
  R.~Lachaize, and M.~Roth.
\newblock {Challenges of memory management on modern NUMA systems}.
\newblock {\em Communications of the ACM}, 58(12):59--66, 2015.

\bibitem{gepner2006multi}
P.~Gepner and M.~F. Kowalik.
\newblock Multi-core processors: New way to achieve high system performance.
\newblock In {\em International Symposium on Parallel Computing in Electrical
  Engineering (PARELEC'06)}, pages 9--13. IEEE, 2006.

\bibitem{ghemawat2015tcmalloc}
S.~Ghemawat and P.~Menage.
\newblock Tcmalloc: Thread-caching malloc.
\newblock \url{github.com/gperftools/}, 2015.

\bibitem{giceva2019os}
J.~Giceva.
\newblock {Operating Systems Support for Data Management on Modern Hardware}.
\newblock \url{sites.computer.org/debull/A19mar/p36.pdf}, 2019.

\bibitem{giceva2014deployment}
J.~Giceva, G.~Alonso, T.~Roscoe, and T.~Harris.
\newblock Deployment of query plans on multicores.
\newblock {\em Proceedings of the VLDB Endowment}, 8(3):233--244, 2014.

\bibitem{giceva2012towards}
J.~Giceva, A.~Sch{\"u}pbach, G.~Alonso, and T.~Roscoe.
\newblock Towards database/operating system co-design.
\newblock In {\em Proceedings of the 2nd workshop on Systems for Future
  Multi-core Architectures (April 2012), SFMA}, volume~12. Citeseer, 2012.

\bibitem{giceva2016customized}
J.~Giceva, G.~Zellweger, G.~Alonso, and T.~Rosco.
\newblock {Customized OS support for data-processing}.
\newblock In {\em Proceedings of the 12th International Workshop on Data
  Management on New Hardware (DaMoN)}, page~2. ACM, 2016.

\bibitem{gray1994quickly}
J.~Gray et~al.
\newblock Quickly generating billion-record synthetic databases.
\newblock {\em Sigmod Record}, 23(2), 1994.

\bibitem{gray2014non}
W.~S. Gray.
\newblock {Non-uniform memory access (NUMA) resource assignment and
  re-evaluation}, Aug.~21 2014.
\newblock US Patent App. 14/178,810.

\bibitem{likwid2010}
G.~Hager, G.~Wellein, and J.~Treibig.
\newblock {LIKWID: A Lightweight Performance-Oriented Tool Suite for x86
  Multicore Environments}.
\newblock In {\em International Conference on Parallel Processing Workshops},
  pages 207--216. IEEE Computer Society, 2010.

\bibitem{hudson2006mcrt}
R.~L. Hudson, B.~Saha, A.-R. Adl-Tabatabai, and B.~C. Hertzberg.
\newblock Mcrt-malloc: a scalable transactional memory allocator.
\newblock In {\em International symposium on Memory management}, pages 74--83.
  ACM, 2006.

\bibitem{mongodb2019}
M.~Inc.
\newblock {The MongoDB 4.0 Manual}.
\newblock \url{docs.mongodb.com/manual/tutorial/transparent-huge-pages/}, 2019.

\bibitem{kaestle2015shoal}
S.~Kaestle, R.~Achermann, T.~Roscoe, and T.~Harris.
\newblock {Shoal: Smart Allocation and Replication of Memory For Parallel
  Programs}.
\newblock In {\em {USENIX} Annual Technical Conference ({USENIX} {ATC} 15)},
  pages 263--276, Santa Clara, CA, 2015. {USENIX} Association.

\bibitem{kambatla2014trends}
K.~Kambatla, G.~Kollias, V.~Kumar, and A.~Grama.
\newblock Trends in big data analytics.
\newblock {\em Journal of Parallel and Distributed Computing},
  74(7):2561--2573, 2014.

\bibitem{karpuzcu2009bubblewrap}
U.~R. Karpuzcu, B.~Greskamp, and J.~Torrellas.
\newblock The bubblewrap many-core: popping cores for sequential acceleration.
\newblock In {\em 2009 42nd Annual IEEE/ACM International Symposium on
  Microarchitecture (MICRO)}, pages 447--458. IEEE, 2009.

\bibitem{kemper2011hyper}
A.~Kemper and T.~Neumann.
\newblock {HyPer: A hybrid OLTP\&OLAP main memory database system based on
  virtual memory snapshots}.
\newblock In {\em IEEE International Conference on Data Engineering (ICDE)},
  pages 195--206. IEEE, 2011.

\bibitem{kiefer2013experimental}
T.~Kiefer, B.~Schlegel, and W.~Lehner.
\newblock Experimental evaluation of numa effects on database management
  systems.
\newblock {\em Datenbanksysteme f{\"u}r Business, Technologie und Web (BTW)
  2025}, 2013.

\bibitem{kim2011multicore}
W.~Kim and M.~Voss.
\newblock Multicore desktop programming with intel threading building blocks.
\newblock {\em IEEE software}, 28(1), 2011.

\bibitem{kissinger2014eris}
T.~Kissinger, T.~Kiefer, B.~Schlegel, D.~Habich, D.~Molka, and W.~Lehner.
\newblock {ERIS: A NUMA-aware in-memory storage engine for analytical
  workloads}.
\newblock {\em Proceedings of the VLDB Endowment}, 7(14):1--12, 2014.

\bibitem{kukanov2007foundations}
A.~Kukanov and M.~J. Voss.
\newblock {The Foundations for Scalable Multi-core Software in Intel Threading
  Building Blocks.}
\newblock {\em Intel Technology Journal}, 11(4), 2007.

\bibitem{kuszmaul2015supermalloc}
B.~C. Kuszmaul.
\newblock {SuperMalloc: a super fast multithreaded malloc for 64-bit machines}.
\newblock In {\em ACM SIGPLAN Notices}, volume~50, pages 41--55. ACM, 2015.

\bibitem{kwon2016coordinated}
Y.~Kwon, H.~Yu, S.~Peter, C.~J. Rossbach, and E.~Witchel.
\newblock Coordinated and efficient huge page management with ingens.
\newblock In {\em {USENIX} Symposium on Operating Systems Design and
  Implementation ({OSDI} 16)}, pages 705--721. {USENIX} Association, 2016.

\bibitem{LangLA0K13}
H.~Lang et~al.
\newblock {Massively Parallel NUMA-Aware Hash Joins}.
\newblock In {\em IMDM}, 2013.

\bibitem{dlmalloc2019}
D.~Lea.
\newblock Dong lea's malloc (dlmalloc).
\newblock \url{gee.cs.oswego.edu/dl/html/malloc.html}, 2000.

\bibitem{leis2014morsel}
V.~Leis et~al.
\newblock {Morsel-driven parallelism: a NUMA-aware query evaluation framework
  for the many-core age}.
\newblock In {\em SIGMOD}. ACM, 2014.

\bibitem{leis2016art}
V.~Leis, F.~Scheibner, A.~Kemper, and T.~Neumann.
\newblock {The ART of practical synchronization}.
\newblock In {\em DaMoN}, pages 1--8. ACM, 2016.

\bibitem{lepers2015thread}
B.~Lepers et~al.
\newblock {Thread and Memory Placement on NUMA Systems: Asymmetry Matters.}
\newblock In {\em USENIX ATC}, 2015.

\bibitem{li2014algorithmic}
X.~Li et~al.
\newblock Algorithmic improvements for fast concurrent cuckoo hashing.
\newblock In {\em EuroSys}. ACM, 2014.

\bibitem{mao2012cache}
Y.~Mao, E.~Kohler, and R.~T. Morris.
\newblock Cache craftiness for fast multicore key-value storage.
\newblock In {\em Proceedings of the 7th ACM european conference on Computer
  Systems}, pages 183--196. ACM, 2012.

\bibitem{mckee2004reflections}
S.~A. McKee et~al.
\newblock Reflections on the memory wall.
\newblock In {\em Conf. Computing Frontiers}, page 162, 2004.

\bibitem{memarzia2019six}
P.~Memarzia, S.~Ray, and V.~C. Bhavsar.
\newblock {A Six-dimensional Analysis of In-memory Aggregation}.
\newblock In {\em International Conference on Extending Database Technology
  (EDBT)}, pages 289--300, 2019.

\bibitem{molka2015cache}
D.~Molka, D.~Hackenberg, R.~Sch{\"o}ne, and W.~E. Nagel.
\newblock Cache coherence protocol and memory performance of the intel
  haswell-ep architecture.
\newblock In {\em International Conference on Parallel Processing (ICPP)},
  pages 739--748. IEEE, 2015.

\bibitem{monetdb2018}
{MonetDB B.V.}
\newblock {MonetDB}.
\newblock \url{monetdb.org}, 2018.

\bibitem{olivier2012openmp}
S.~L. Olivier et~al.
\newblock Openmp task scheduling strategies for multicore numa systems.
\newblock {\em IJHPCA}, 26(2), 2012.

\bibitem{oracle2019}
Oracle.
\newblock {Grid Infrastructure Installation and Upgrade Guide for Linux}.
\newblock
  \url{docs.oracle.com/en/database/oracle/oracle-database/12.2/cwlin/disabling-transparent-hugepages.html},
  2019.

\bibitem{Porobic14ATraPos}
D.~Porobic, E.~Liarou, P.~Tozun, and A.~Ailamaki.
\newblock Atrapos: Adaptive transaction processing on hardware islands.
\newblock pages 688--699, 03 2014.

\bibitem{powers1998applications}
D.~M. Powers.
\newblock {Applications and explanations of Zipf's law}.
\newblock In {\em NeMLaP3/CoNLL98}, pages 151--160. Association for
  Computational Linguistics, 1998.

\bibitem{psaroudakis2015scaling}
I.~Psaroudakis et~al.
\newblock {Scaling up concurrent main-memory column-store scans: towards
  adaptive NUMA-aware data and task placement}.
\newblock {\em VLDBJ}, 8(12), 2015.

\bibitem{psaroudakis2018analytics}
I.~Psaroudakis, S.~Kaestle, M.~Grimmer, D.~Goodman, J.-P. Lozi, and T.~Harris.
\newblock Analytics with smart arrays: adaptive and efficient
  language-independent data.
\newblock In {\em Proceedings of the Thirteenth EuroSys Conference}, page~17.
  ACM, 2018.

\bibitem{Psaroudakis2016AdaptiveNUMA}
I.~Psaroudakis, T.~Scheuer, N.~May, A.~Sellami, and A.~Ailamaki.
\newblock {Adaptive NUMA-aware Data Placement and Task Scheduling for
  Analytical Workloads in Main-memory Column-stores}.
\newblock {\em Proc. VLDB Endow.}, 10(2):37--48, Oct. 2016.

\bibitem{pugh1990skip}
W.~Pugh.
\newblock Skip lists: a probabilistic alternative to balanced trees.
\newblock {\em Communications of the ACM}, 33(6), 1990.

\bibitem{redhat2018}
{Red Hat Inc}.
\newblock {Red Hat Enterprise Linux Product Documentation}, 2018.

\bibitem{redhat2018thp}
{Red Hat Inc}.
\newblock {Red Hat Performance Tuning Guide}.
\newblock
  \url{access.redhat.com/documentation/en-us/red_hat_enterprise_linux/6/html/performance_tuning_guide/s-memory-transhuge},
  2018.

\bibitem{redis2019}
Redis.
\newblock Redis latency problems troubleshooting.
\newblock \url{redis.io/topics/latency}, 2019.

\bibitem{rogers2009scaling}
B.~M. Rogers, A.~Krishna, G.~B. Bell, K.~Vu, X.~Jiang, and Y.~Solihin.
\newblock Scaling the bandwidth wall: challenges in and avenues for cmp
  scaling.
\newblock {\em ACM SIGARCH Computer Architecture News}, 37(3):371--382, 2009.

\bibitem{schuh2016experimental}
S.~Schuh et~al.
\newblock An experimental comparison of thirteen relational equi-joins in main
  memory.
\newblock In {\em SIGMOD}, 2016.

\bibitem{singh20173}
T.~Singh, S.~Rangarajan, D.~John, C.~Henrion, S.~Southard, H.~McIntyre,
  A.~Novak, S.~Kosonocky, R.~Jotwani, A.~Schaefer, et~al.
\newblock 3.2 zen: A next-generation high-performance$\times$ 86 core.
\newblock In {\em IEEE International Solid-State Circuits Conference (ISSCC)},
  pages 52--53. IEEE, 2017.

\bibitem{sivarajah2017critical}
U.~Sivarajah, M.~M. Kamal, Z.~Irani, and V.~Weerakkody.
\newblock Critical analysis of big data challenges and analytical methods.
\newblock {\em Journal of Business Research}, 70:263--286, 2017.

\bibitem{srikanthan2015data}
S.~Srikanthan, S.~Dwarkadas, and K.~Shen.
\newblock Data sharing or resource contention: Toward performance transparency
  on multicore systems.
\newblock In {\em {USENIX} Annual Technical Conference ({USENIX} {ATC} 15)},
  pages 529--540. {USENIX} Association, 2015.

\bibitem{stonebraker2013voltdb}
M.~Stonebraker and A.~Weisberg.
\newblock {The VoltDB Main Memory DBMS}.
\newblock {\em IEEE Data Eng. Bull.}, 36(2):21--27, 2013.

\bibitem{gcc2019}
G.~Team et~al.
\newblock {GCC, the gnu compiler collection}.
\newblock \url{gcc.gnu.org}, 2019.

\bibitem{glibc2019}
{The glibc project developers}.
\newblock {The GNU C Library (glibc)}.
\newblock \url{gnu.org/software/libc/}, 2019.

\bibitem{postgres2019}
{The PostgreSQL Global Development Group}.
\newblock {PostgreSQL}.
\newblock \url{postgresql.org}, 2019.

\bibitem{umayabara2017mcmalloc}
A.~Umayabara and H.~Yamana.
\newblock {MCMalloc: A scalable memory allocator for multithreaded applications
  on a many-core shared-memory machine}.
\newblock In {\em 2017 IEEE International Conference on Big Data (Big Data)},
  pages 4846--4848. IEEE, 2017.

\bibitem{van2017memory}
H.~Van~Riel.
\newblock Memory distribution across multiple non-uniform memory access nodes,
  Oct.~10 2017.
\newblock US Patent 9,785,581.

\bibitem{villars2011big}
R.~L. Villars, C.~W. Olofson, and M.~Eastwood.
\newblock Big data: What it is and why you should care.
\newblock {\em White Paper, IDC}, 14:1--14, 2011.

\bibitem{skiplist2016}
S.~Vokes.
\newblock {skiplist}.
\newblock \url{github.com/silentbicycle/skiplist}, 2016.

\bibitem{wang2015numa}
L.~Wang et~al.
\newblock {NUMA-aware scalable and efficient in-memory aggregation on large
  domains}.
\newblock {\em TKDE}, 27(4), 2015.

\bibitem{wang2016predicting}
W.~Wang, J.~W. Davidson, and M.~L. Soffa.
\newblock {Predicting the memory bandwidth and optimal core allocations for
  multi-threaded applications on large-scale NUMA machines}.
\newblock In {\em IEEE International Symposium on High Performance Computer
  Architecture (HPCA)}, pages 419--431. IEEE, 2016.

\bibitem{Wentzlaff:2009:FOS}
D.~Wentzlaff and A.~Agarwal.
\newblock Factored operating systems (fos): The case for a scalable operating
  system for multicores.
\newblock {\em SIGOPS Oper. Syst. Rev.}, 43(2):76--85, Apr. 2009.

\end{thebibliography}
\nocite{*}

\end{document}